\documentclass[twocolumn]{aastex61}
\usepackage{comment}

\newcommand{\mgediez}{\log{(\rm M_{\star}/M_{\odot})}\ge10}
\newcommand{\mltdiez}{\log{(\rm M_{\star}/M_{\odot})}<10}
\submitjournal{ApJ}

\shorttitle{Star Formation Properties in $z<2.5$ Mergers}
\shortauthors{Silva et al.}
\begin{document}
\title{Galaxy mergers up to $z<2.5$ I :  The star formation properties of merging galaxies at separations 3-15 kpc}
\correspondingauthor{Silva et al.}
\email{andrea.silva@nao.ac.jp}

\author{Andrea Silva}
\affil{Department of Physics and Astronomy, Tufts University, Medford, MA 02155, USA}
\affil{National Astronomical Observatory of Japan, National Institutes of Natural Sciences, 2-21-1 Osawa, Mitaka, Tokyo 181-8588, Japan}

\author{Danilo Marchesini}
\affil{Department of Physics and Astronomy, Tufts University, Medford, MA 02155, USA}

\author{John D. Silverman}
\affil{Kavli Institute for the Physics and Mathematics of the Universe (WPI), The University of Tokyo Institutes for Advanced Study, The University of Tokyo, Kashiwa, Chiba 277-8583, Japan}

\author{Rosalind Skelton}
\affil{South African Astronomical Observatory, PO Box 9, Observatory, Cape Town 7935, South Africa}

\author{Daisuke Iono}
\affil{National Astronomical Observatory of Japan, National Institutes of Natural Sciences, 2-21-1 Osawa, Mitaka, Tokyo 181-8588, Japan}

\author{Nicholas Martis}
\affil{Department of Physics and Astronomy, Tufts University, Medford, MA 02155, USA}

\author{Z. Cemile Marsan}
\affil{Department of Physics and Astronomy, York University, 4700 Keele Street, Toronto, Ontario, MJ3 1P3, Canada}

\author{Ken-ichi Tadaki}
\affil{National Astronomical Observatory of Japan, National Institutes of Natural Sciences, 2-21-1 Osawa, Mitaka, Tokyo 181-8588, Japan}

\author{Gabriel Brammer}
\affil{Space Telescope Science Institute, 3700 San Martin Drive, Baltimore, MD 21218, USA}

\author{Jeyhan kartaltepe}
\affil{School of Physics and Astronomy, Rochester Institute of Technology, 84 Lomb Memorial Drive, Rochester, NY 14623, USA}

\begin{abstract}
We present a study of the influence of galaxy mergers on star formation 
at $0.3<z<2.5$.
Major mergers are selected from the CANDELS/3D-HST catalog using a peak-finding algorithm. Mergers have projected galaxy nuclei separation of their members 
between 3-15 kpc.
 We compare the star formation activity in merging and  non-merging galaxies and find no significant differences. We find that  only 12\% of the galaxies in major mergers (in which both galaxies have $\mgediez$) are star-bursting (i.e., with SFR above the main sequence of star-forming galaxies by $>$0.5 dex).
    Merging galaxies which include galaxies with  lower masses show a higher fraction of star-bursting  galaxies (20\%).
The low fraction of star-bursting merging galaxies in this sample suggests that  at galaxy nuclei separations of 3-15 kpc merging galaxies are still in a early stage and are yet to reach the maximum level of star formation activity.  
Furthermore, the level of star formation enhancement and its duration could be arguably reduced compared to local mergers, as shown by simulations of high-$z$ mergers, and might also depend on the  physical properties (such as stellar mass and gas fraction) of the merging galaxies. 
 Finally, we compare the specific SFR between  merging galaxies.  Our results suggest that, as the mass of the merging galaxies increases, the star formation activity in the less massive member in the merger suffers a more dramatic impact than its companion galaxy.

\end{abstract}

\section{Introduction} \label{sec_intro}

  The complete path of how baryonic matter evolves to form the galaxies we see today is still unclear.
   In the $\Lambda$CDM picture  \citep{riess1998, perlmutter1999, parkinson2012}, galaxies and their host dark matter halos grow via the merging  of increasingly more massive dark matter halos  \citep{cole2008, neistein2008}. 
However, this hierarchical growth is not the only channel to build galaxies.  
Galaxies can also be fed by  cold gas in filamentary streams \citep{dekel2009}.  In addition, observations show that low-mass galaxies formed most of their stars at later times than more massive systems  \citep[so called ``down-sizing",][]{white1991, kauffmann1993, kauffmann1999} contrary to the hierarchical assembly.  
Therefore, to understand the importance of the hierarchical growth, it is necessary to study the role of galaxy mergers  in the formation and evolution of galaxies.
Specifically, the effects of merging on the star formation activity and on the morphological transformation of the merging systems at different cosmic times are yet to be robustly assessed. 
Similarly, the role played by mergers in feeding and growing the central super-massive black holes is far from being understood, especially at early cosmic times. 
This is  crucial as the energetic output from the active galactic nuclei (AGN) is thought to contribute to, if not dominate, the quenching of star formation in galaxies (AGN feedback) in massive halos.

Interactions of galaxies with similar mass (major mergers, mass ratio $\ge$ 1:4) have been studied in depth in the local Universe.  Major mergers 
can enhance  star formation in the system if the members have sufficient gas  \citep{kartaltepe2012, yuan2012, hung2013, patton2013, lanz2013}. 
The interaction can also morphologically transform galaxies and trigger AGN
\citep{barnes1996, mihos1996, springel2005, hopkins2006, hopkins2008}.
 Observations have shown that almost all  local starbursts, which are galaxies that are located well above the main sequence, are produced by major mergers 
\citep{armus1987, sanders1996, kartaltepe2010,ellison2013}.  
However, starbursts may not necessarily be produced by mergers at high redshifts, because the physical conditions in the Universe were different, such as the availability of more gas to produce stars even if not in a merger  \citep{daddi2010, tacconi2010}.

 Observations of merging systems at high-$z$ have shown that major mergers are inefficient at driving star formation \citep{kaviraj2013, kaviraj2015, lofthouse2017}.
 However, these studies have not  established how the process of driving star formation  depends on the 
  different merger stages.  Hints of this process have been shown by simulations of galaxy mergers.
  These simulations assume that high-$z$ merging galaxies contain higher gas fractions than local mergers.  
  The results show a  weak increment  in the intensity and duration of  the star formation enhancement  produced by the high-$z$ mergers  \citep{perret2014, scudder2015, fensch2017}.
In the case of major mergers of two gas poor galaxies (dry mergers), there is little or no star formation enhancement in the system, although the interaction contributes to the build up of massive galaxies \citep{cattaneo2008}.   The dry merger scenario tends to be more common at low redshift   \citep{khochfar2003}. 

Observationally,  several techniques have been used to select galaxy mergers to quantify the role played by mergers in galaxy formation and evolution. 
First, selecting pairs of galaxies with projected separations $\lesssim$100$h_{100}^{-1}$ kpc \citep[e.g.,][]{robaina2010, williams2011, tasca2014} is a common method of identifying candidate mergers before coalescence 
 takes place.
 One issue with this method is that it includes 
contamination by chance superpositions if redshift information  is not available.
A second technique is based on the selection of 
 morphologically disturbed galaxies with signatures such as asymmetric features,
 tidal tails, or  outer shells 
that  are sensitive to minor mergers and close passages  \citep[e.g.,][]{bridge2010, cisternas2011, kartaltepe2015}.

A third technique was introduced by \citet{lackner2014}. It   selects merging galaxies at the interface between early-stage mergers selected in close galaxy pairs studies and post-merger galaxies based on disturbed morphologies. 
The technique uses 
a high-pass filter that selects bright peaks in a galaxy surface brightness map (i.e., double nuclei).  \citet{lackner2014} applied  this method to the ACS I$_{814}$-band images in the full COSMOS field.  The  I$_{814}$-band limited the analysis to $z\lesssim 1$  to probe the rest-frame optical redward of the Balmer 4000\AA~ break.  
\citet{lackner2014} selected galaxies with two intact nuclei separated by 2.2-8 kpc which are expected to merge within a few hundreds Myrs.
This technique has two main advantages over past efforts. First, the method can be applied to photometric samples that lack spectroscopy due to the small separation of the nuclei thus the sample sizes are greatly increased. Second, the selection function can be quantified by constructing simulated mergers using isolated (real) galaxies and determining the success rate with respect to recovering the detection of both peaks in the light distribution. The latter is important when measuring the redshift dependence of the merger rate. 
Since \citet{lackner2014} used only one HST band in their study, they could not distinguish the colors of 
the individual member galaxies in mergers. 
The lack of resolved information prevented the estimation of the stellar population properties of  member galaxies, and 
 their mass ratios were not measured to differentiate
 major from minor mergers.\footnote{They selected major from minor mergers from the flux ratios of the merger members. The fluxes were calculated from the brightest regions in the galaxies (the nuclei).}
 They also could not distinguish  between wet, mixed, and dry mergers based on the individual colors of the merging galaxies and their sample is potentially affected by chance superpositions (due to the lack of resolved redshifts) and clumpy galaxies.

 To make further progress at $z>1$ using the \citet{lackner2014} technique,  it is necessary to apply it to NIR data
to obtain a sample of mergers less contaminated by clumpy, star-forming galaxies (which are more likely to be falsely identified as merging systems in bluer bands).
This technique was shown to be more sensitive to mergers between concentrated galaxies  of early-type morphology and is therefore  ideally suited to longer wavelength data in which galaxies appear more bulge-dominated and centrally concentrated. 

In this work, we use the imaging and spectroscopic data of the CANDELS and 3D-HST datasets to study the resolved star-formation properties of merging galaxies with projected separation of 3-15 kpc, over the redshift range 0.3$<z<$2.5. 
With HST data from the optical to the NIR, we extract  spatially resolved spectral energy distributions (SEDs), colors, and accurate redshifts up to $z$=2.5. 
This resolved information is crucial  for a detailed study of the properties of the merging galaxies.
  Mergers are selected using the technique introduced by \citet{lackner2014}  on the HST F160W (H160 hereafter) images. The multi-wavelength space-based imaging from CANDELS and the grism spectroscopic redshifts from 3D-HST are used to remove chance superposition of galaxies, clean the sample from clumpy galaxies, confirm interacting pairs, as well as to study the resolved star-formation properties of the merging galaxies.

In section \S  \ref{sec_data} we present the data and describe the procedure to select mergers.   
In section \S \ref{sec_results} we present the  star formation properties of merging galaxies and compare them with the properties of non-merging galaxies. 
 In \S \ref{sec_discussion} and \S \ref{sec_conclusion} we present the discussions and conclusions.
In future papers, we will present the merger rate, structural properties of these merging galaxies, and the incidence of AGN. 
Throughout this paper, we adopt a cosmology with H$_{0}$=70 km s$^{-1}$ Mpc$^{-1}$, $\Omega_{\Lambda}$=0.7, and $\Omega_{m}$=0.3.  Magnitudes are in the AB system.

\section{Data \& Sample} \label{sec_data}

\subsection{3D-HST Photometric Catalog and Grism Redshifts}\label{sec_catalog}

We use the 3D-HST  \citep{skelton2014, momcheva2016} catalogs and images to find galaxy pairs. The catalog covers all five CANDELS fields (COSMOS, AEGIS, GOODS-North, GOODS-South, and UDS).  We make use of the derived data products produced by the 3D-HST collaboration in these fields, i.e. photometric redshift estimates, and stellar and structural parameters such as stellar masses, ages, star formation rates, extinction, and sizes. 
The catalogs contain total flux measurements and stellar population parameters for 207,967 objects over an effective area of 896.3 arcmin$^{2}$.  
 G141 grism redshifts for 22,548 galaxies are provided by \citet{momcheva2016}. We use the best redshift available for each galaxy (spectroscopic, grism, or photometric redshift). 
 Stellar masses were calculated by fitting the galaxy SEDs using  \citet{bruzual2003}
 stellar population models using FAST \citep{kriek2009}, with \citet{chabrier2003} initial mass function (IMF), an exponentially declining star formation history\footnote{We have tested the robustness of the stellar mass and SFR estimates by also adopting a delayed exponentially declining star formation history (which allows for initially SFR that increases with time). Stellar mass estimates are very robust, with no systematic difference and a negligible scatter of 0.05 dex. SFR are also quite robust, with a small systematic difference $<10\%$ and a scatter of $\sim$0.2 dex.}, and a solar metallicity.
  The star formation rate was obtained from the combination of  rest-frame UV emission and the mid-infrared photometry obtained from {\it Spitzer}/MIPS imaging following \citet{whitaker2012}.
  When MIPS 24 $\mu$m data were not available or for detections with S/N$<$3, the star formation rates were obtained from the modeling  of the SEDs. 

We create postage stamps for each galaxy from the near-infrared H160 images. The stamps have a size of 8\arcsec $\times$ 8\arcsec.  The pixel size of these images is 0\farcs06 and the point spread function has a FWHM of 0\farcs19.  
We remove stars ({\sc star\_flag}$\neq$1) and objects too close to bright stars ({\sc near\_star}=0) from the catalog. We select sources with enough exposure time in F125W and F160W bands ({\sc nexp\_125W}, {\sc nexp\_160W}$\ge$2.0) and with signal to noise ratio in F160W S/N$>$10.

\vspace{2cm}
\subsection{Detection of Galaxy Pairs}\label{sec_detection}

We select galaxy pairs\footnote{Galaxy pairs: two galaxies at close projected nuclei separation that are not necessarily merging galaxies} by applying the technique described in  \citet{lackner2014} to  the H160 postage stamps.
 The method consists of selecting bright peaks in an image and then implementing
 restrictions on the properties of the peaks  to select galaxy pairs.
 We apply the technique 
  to 5717 galaxies in the 3D-HST catalog with $\mgediez$, brighter than $m_{\rm AB}$=24.5 in the H160 band, and $0.3<z<2.5$\footnote{ We use a minimum redshift of $z=0.3$ because at this redshift two galaxy pairs at a distance of 15 kpc are at a projected separation of 3\farcs4, therefore both galaxies fall within the postage stamp.  The maximum redshift of  $z=2.5$ is used because  the completeness in stellar mass at $\mgediez$ is above 90\%. }.
 The postage stamps are centered at the position of these galaxies and are convolved with a median ring filter that smooths the image. 
  This  erases structures on scales larger than the ring. 
The optimal ring size to select bright peaks in the H160 postage stamps has a diameter of  2$\times$FWHM (0\farcs38).
 The size of the ring corresponds to 1.7 and 3.1 kpc at $z$=0.3 and $z$=2.5, 
 respectively. 
We subtract the smoothed from the original image and from the resultant image, we select all the regions that contain at least 9 pixels with signal to noise ratio $> 5\sigma$. 
These regions correspond to bright peaks in the image. The fluxes of these peaks are obtained by summing up the pixel values associated with the peak in the original image.
 To find galaxy pairs with similar projected separation at different redshifts, we choose peaks that have separation between 3 to 15 kpc\footnote{Assuming that the system is at the redshift of the central galaxy.}. The minimum separation is set by the size of the median ring filter at $z$=2.5, and the upper separation is to avoid line of sight contaminating galaxies. 
We are interested in finding galaxy pairs that potentially are major merging galaxies. 
Following \citet{lackner2014}, we select peaks that have at least 1:4 of the flux of the brightest peak.
This restriction also helps to remove star-forming clumps that could contaminate the sample of selected galaxy pairs. 
 To avoid edge on disk galaxies, in which 3 peaks are aligned, we set that the Pearson correlation coefficient between the peaks is less than 0.5.  
The number of galaxy pairs obtained with these restrictions is 678.

\subsection{Blended Sources in the 3D-HST Catalog}\label{sec_blended}

\begin{figure*}[!htbp]
\begin{center}
\includegraphics[angle=0,scale=0.60]{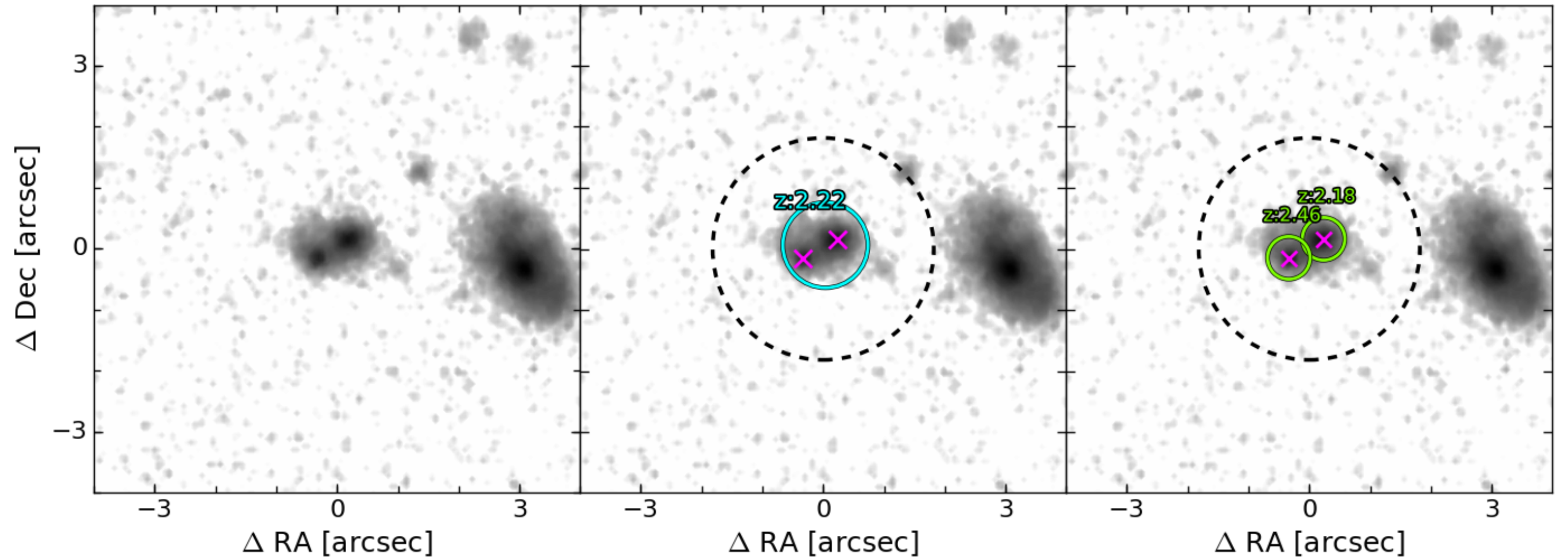}
\caption{Example of a galaxy pair identified using the peak-finding algorithm which is blended (not resolved) in the 3D-HST catalog. {\bf Left:} Postage stamp centered at the position of a source identified by the 3D-HST catalog.   {\bf Middle:} 
The cyan circle
is centered at the position of the source identified in the 3D-HST catalog and  with a size equal to the  aperture used to calculate photometry (D=0\farcs7). The measured redshift by the 3D-HST team  is indicated. The dashed circle shows the region where we search for galaxy pairs (15 kpc in radius at the redshift of the central source).
Magenta crosses indicate the position of the bright peaks identified with the peak-finding code.    {\bf Right:}  Green circles show the position and size of the apertures used to calculate de-blended photometry using an aperture of D=0\farcs3. Their resolved photometric redshifts are indicated.  
\label{fig_blended}}
\end{center}
\end{figure*}

In the sample of 678 galaxy pairs found with the peak-finding code, 28\% of the galaxies are blended (not resolved) in the 3D-HST catalog (Fig. \ref{fig_blended}). 
We extract de-blended HST photometry for the individual   blended member galaxies using the same procedure performed to create the 3D-HST photometric catalogs, except that, for the not resolved sources, we use an aperture of 0\farcs3 in diameter (instead of an aperture of 0\farcs7 as used for the 3D-HST catalogs). 
The H160 stamps of the blended sources are modeled with GALFIT \citep{peng2002} to obtain the H160 total fluxes of the blended galaxies. The total H160 fluxes are used to scale the D=0\farcs3 aperture photometry to the total  by multiplying by the ratio  $f_{\rm total}/f_{\rm D=0\farcs3}$ in the H160 band. These SEDs are then modeled to derive photometric redshifts and rest-frame luminosities and colors with EAZY \citep{brammer2008} and stellar population properties with FAST \citep{kriek2009} using the same procedure and SED-modeling assumptions adopted for the 3D-HST catalogs.
After this procedure, we select those pairs in which both galaxy members have $m_{\rm AB}\le$24.5. 
 The peak-finding algorithm is applied to galaxies at the center of the postage stamps with masses $\mgediez$.
Since we apply a flux ratio cut to find galaxy pairs, the companion member of the central source in the pair could have a  lower stellar mass than this limit.  
Moreover, for pairs that were originally blended in the 3D-HST catalog, it is possible that both components end up with stellar masses $\mltdiez$
 if the original blended object had  $10<\log(\rm M_{\star}/\rm M_{\odot}) <10.3$.

\subsection{Final Sample of Mergers}\label{sec_finalsample}

Using the redshift values (spectroscopic, grism, or photometric) of the individual member galaxies, we separate galaxy pairs into likely mergers and line of sight contaminants.  
We define  mergers    as
galaxies that are consistent with being at the same redshift within a 3$\sigma$ uncertainty (when using either photometric or grism redshifts), or if the redshifts differ by less than 0.001 (if both have spectroscopic redshifts). 

\begin{figure*}[!htbp]
\begin{center}
\includegraphics[angle=0,scale=0.38]{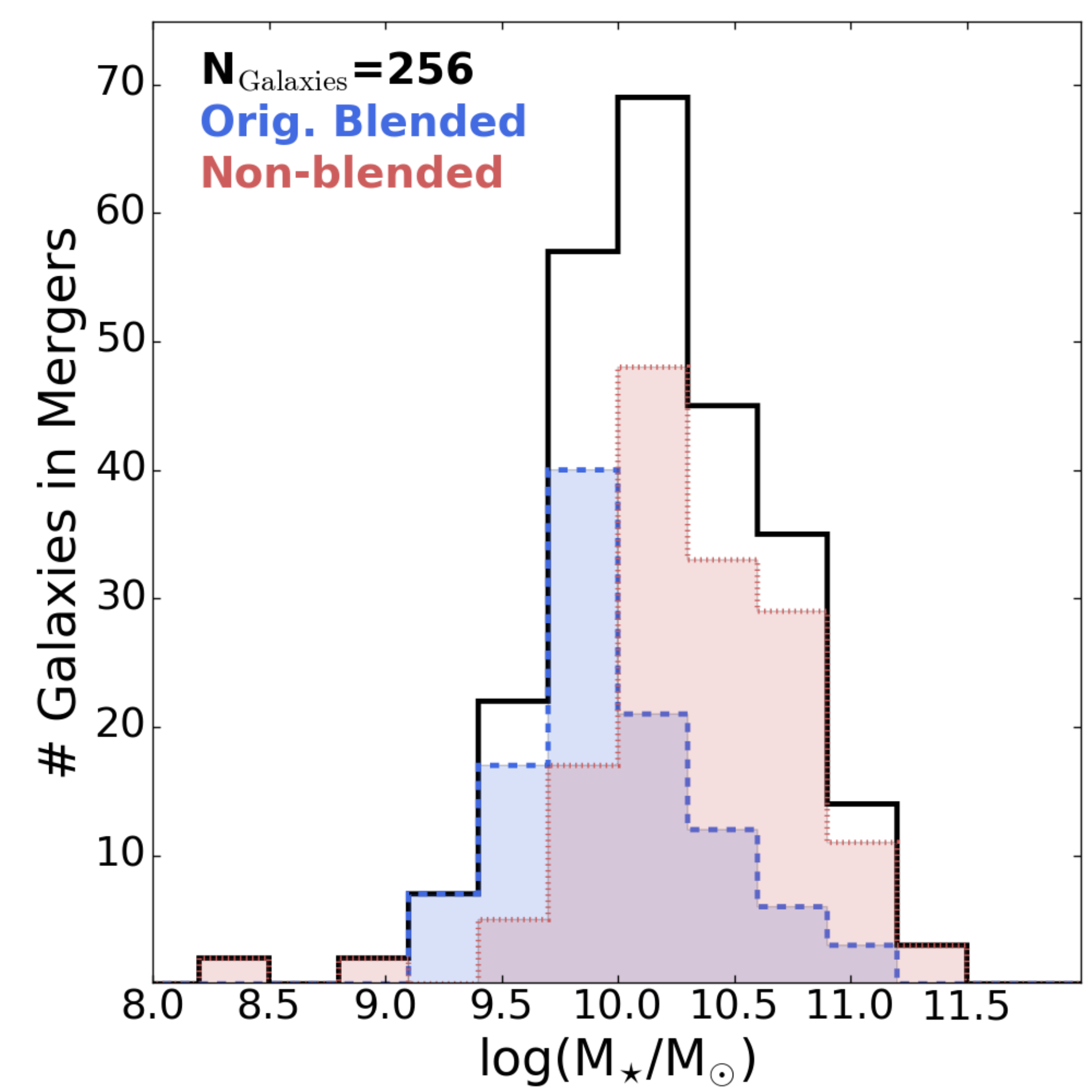}
\includegraphics[angle=0,scale=0.38]{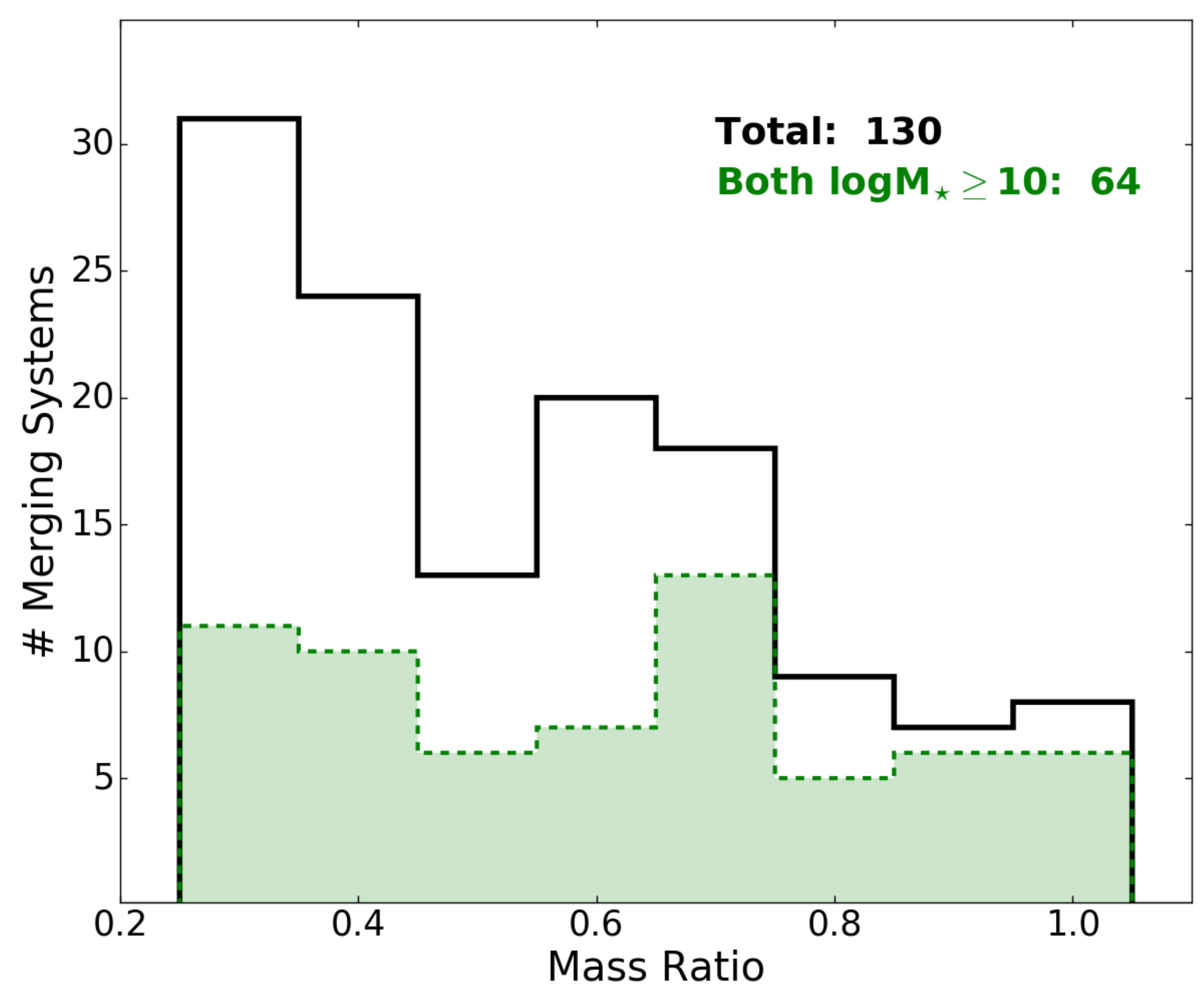}
\caption{{\bf Left}: Distribution of the number of galaxies in major mergers (Section \S \ref{sec_finalsample}) in the primary sample (which include galaxies with masses $\log(\rm M_{\star}/M_{\odot})<10$)  as a function of stellar mass.  
 The black line shows the distribution of all the galaxies in mergers, while the dashed blue and dotted red lines show the distribution of galaxies that were originally  blended (now de-blended) and non-blended in 3D-HST, respectively.   
{\bf Right:} Distribution of merging systems as  function of the mass ratio  of the merger components 
for the primary sample  of mergers (black distribution) and the high-mass sample of  mergers  in which both merging galaxies have masses $\log(\rm M_{\star}/M_{\odot})\ge10$ (green distribution). 
  \label{fig_ngal_m}}
\end{center}
\end{figure*}

As indicated in \S \ref{sec_detection},
 the code selects galaxy pairs with flux ratio$\ge$1:4 to find major mergers. However, the fluxes measured by the code are only a lower limit of the flux of the merging galaxies, since this is measured from the brightest regions in the galaxy.  
 Since we have the stellar masses of the merging galaxies obtained from the 3D-HST catalog (and from the de-blended photometry in originally blended sources), we can 
obtain a sample of major mergers using a cut in mass ratio $\ge$1:4 in addition to the original flux ratio cut\footnote{As presented in Appendix \ref{comparison_ratios}, we do not see a significant difference in the analysis of the properties of mergers when different combination ratio cuts is applied.}. 
The sample of major mergers at $0.3<z<2.5$ in the five CANDELS fields contains 256 merging galaxies in 130 merging systems of which 125 are mergers of two galaxies and the other 5 systems are  mergers with more than two galaxies\footnote{For instance, in a merging system with  three galaxies A,B, and C,  the pairs with galaxies  A-B and A-C but not necessarily B-C could be selected as a merging system (if they follow all the restrictions imposed to select mergers). In this case we count  two merging systems.}. 
 The left panel of  Figure \ref{fig_ngal_m} shows the distribution of masses for  the galaxies in major  mergers.

  The original flux ratio cut helps to remove the selection of clumpy galaxies.  However, star-forming clumps in galaxies 
 usually have stellar masses $8\le \log(\rm M_{\star}/M_{\odot}) \le 10$ \citep{guo2012}  and it might be possible that  the major merger sample could be contaminated by clumpy galaxies. 
  Therefore, we will also analyze a more restrictive sample of major mergers, in which both merging galaxies have stellar masses $\log(\rm M_{\star}/M_{\odot})\ge 10$.
  We will refer to this sample as the ``high-mass" sample of mergers while for the original sample we will refer as the ``primary" sample.   
 The high-mass sample contains  128 galaxies in 64 merging systems (Table \ref{tbl_merger}). 
 We define as non-merging galaxies those galaxies that were not selected in the primary sample. 
 The right panel in Figure \ref{fig_ngal_m} shows the distribution in mass ratio of major mergers, both the primary and the high-mass sample.
 The fraction of galaxies in  the high-mass  sample at $0.3<z<2.5$  is on average  2.7\% with 1.7\% of the mergers at $0.3 < z \le 1.0$ and 2.4\% at $1.0 < z \le 2.5$.

\begin{figure*}[!htbp]
\begin{center}
\includegraphics[angle=0,scale=0.60]{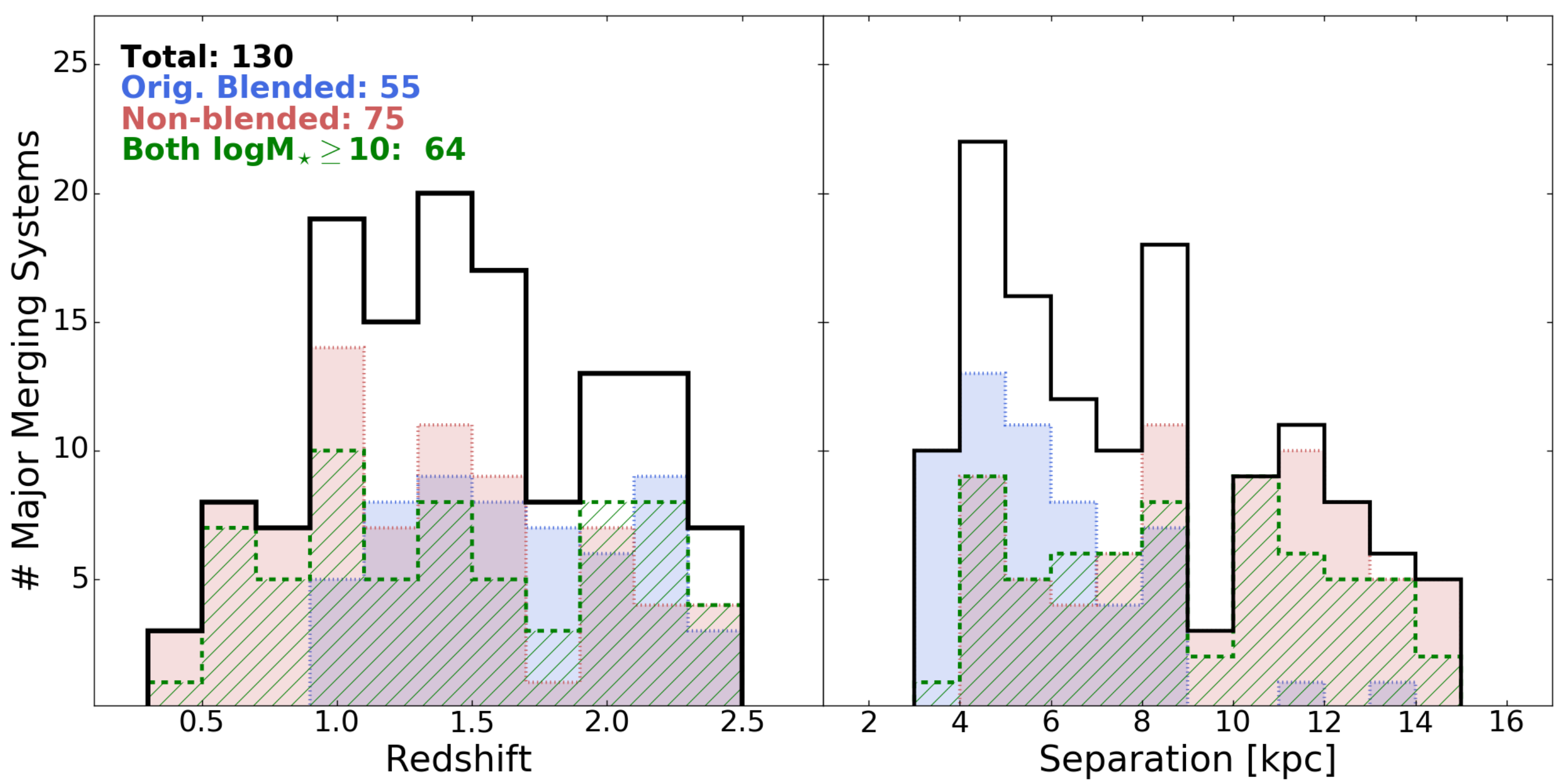}
\caption{Number of major merging systems as a function of redshift (left) and projected separation (right). 
  Black line shows the distribution of the primary sample of merging systems. The blue and red distributions are for the systems in this sample that were originally blended and non-blended in 3D-HST, respectively.  The green distribution is for the high-mass merger sample.   \label{fig_ngal_zbl}}
\end{center}
\end{figure*}

  Figure \ref{fig_ngal_zbl} shows the distribution in redshift and nuclei separation of both major merger samples.  
For the primary sample,  we show the distribution of the systems that were originally blended and those that were resolved in 3D-HST. 
 The combination of the peak-finding algorithm and the de-blending analysis allowed us to increase the sample of major mergers by  73\% compared to the sample of mergers that would have been constructed using the 3D-HST catalogs alone (26\% for the high-mass sample).  This increase is particularly substantial for systems with smaller  projected separations ($\lesssim$9 kpc) and  higher redshifts ($z\gtrsim 1$).
 For the following  analysis in this paper, we will focus on the high-mass major  merger sample  unless explicitly stated otherwise. 

\begin{deluxetable}{lccc   }[!htbp]
\tabletypesize{\scriptsize}
\tablecaption{Number of  Major Merging Systems per Field. \label{tbl_merger} }
\tablewidth{0pt}
\tablehead { Field &  Primary$^{a}$ & High-mass$^{b}$ & Fraction$^{c}$ (\%) } \startdata
 COSMOS            &  24              &   14                      &  2.5  \\  
 AEGIS                 &  27              &   14                      &  2.5  \\
 GOODS-N           &  24              &   8                        & 1.4     \\
 GOODS-S           &  33              &   17                      &   2.9  \\
 UDS                    &  22               &  11                       &  1.7 \\ \hline 
 TOTAL                 &  130            &   64                       &   2.7 \\
  \enddata
 \tablenotetext{a}{ The original sample of mergers obtained using a flux and mass ratio cut $\ge$1:4.}
\tablenotetext{b}{Both merging galaxies have stellar masses $\log(\rm M_{\star}/M_{\odot})\ge 10$.}
\tablenotetext{c}{Fraction of galaxies in the high-mass sample of mergers at $0.3<z<2.5$.} 

\end{deluxetable}


\section{Results} \label{sec_results}

We use the rest-frame U-V and V-J colors to place  the sample of merging galaxies in the U-V versus V-J color-color diagram (UVJ diagram, hereafter).
 In this way, we can  separate merging galaxies into quiescent and star-forming by using the criteria defined  in \citet{whitaker2015}. We also include the criteria presented in  \citet{martis2016} to separate star-forming galaxies into dusty and relatively unobscured star-forming galaxies. 
 Star forming galaxies satisfy 
$(U-V)<1.3$  for $(V-J)<0.75$ and 
$(U-V)< 0.8(V-J)+0.7$ for $(V-J)\ge0.75$,
while dusty galaxies satisfy
$(U-V)<1.43(V-J)-0.36$.

\begin{figure*}[!htbp]
\begin{center}
\includegraphics[angle=0,scale=0.66]{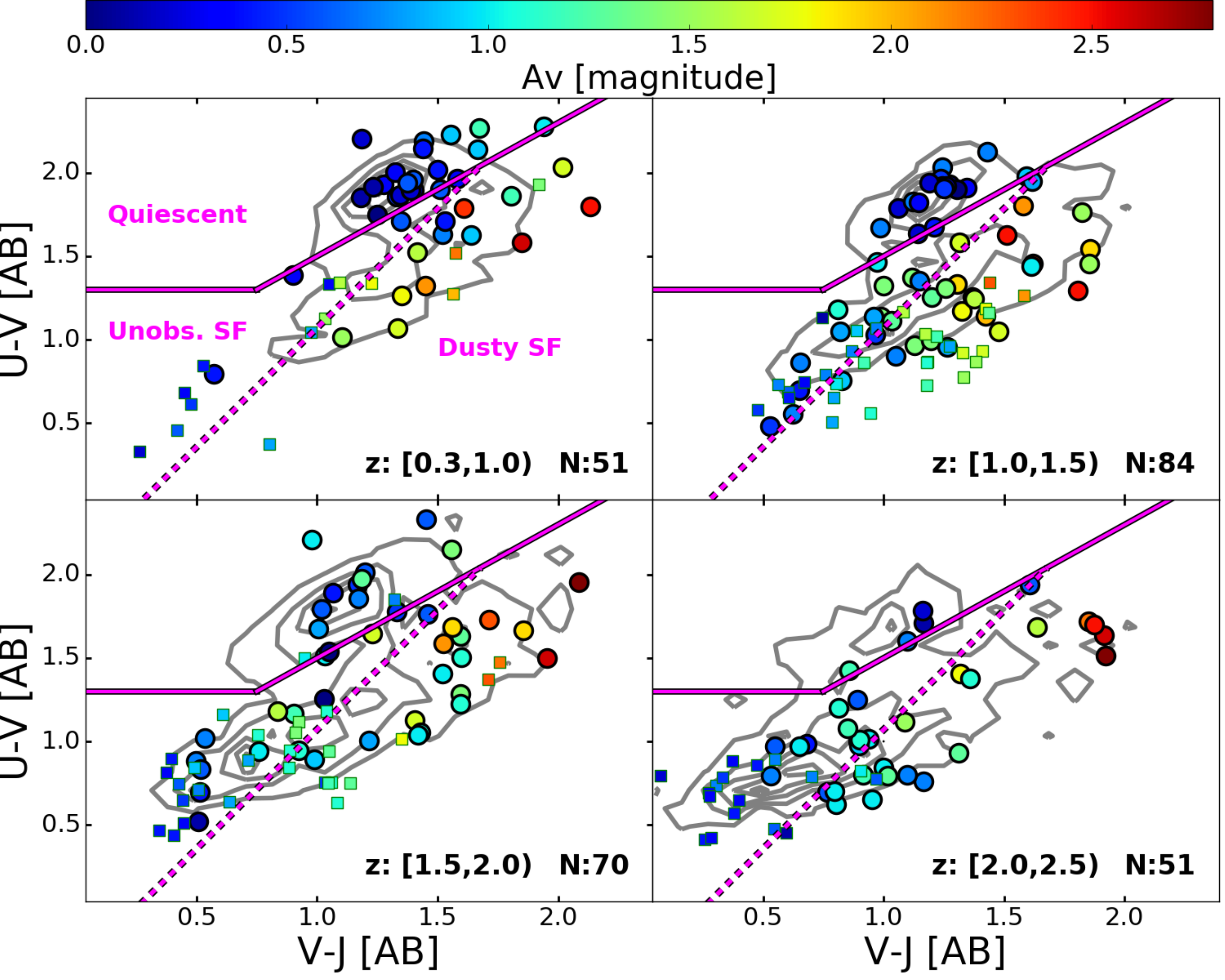}  
\caption{UVJ diagram color coded by Av magnitude for galaxies in major mergers  in different redshift bins.
The gray contours correspond to  non-merging galaxies with $\mgediez$. 
Contours are 10, 30, 50, 70, and 90\% of the distribution.  The solid lines indicate the separation of quiescent (Q)  and star-forming (SF)  galaxies \citep{whitaker2015} and the dashed lines separate  star-forming galaxies into dusty and unobscured \citep{martis2016}. The number of merging galaxies in each redshift bin  is indicated. 
Galaxies with  masses $\mltdiez$ are shown in squares while those with higher masses are indicated with circles.
 \label{fig_uvj}}  
\end{center}
\end{figure*}

Figure \ref{fig_uvj} shows the position of the merging sources (including those of the primary sample) in the UVJ diagram in different redshift bins. We compare their positions with the positions of non-merging galaxies with masses $\mgediez$. 
 For the 128 galaxies in the high-mass major merger sample, we find that 35.9$\pm$5.3\%,  21.9$\pm$3.4\%, and  42.2$\pm$5.7\% are quiescent, unobscured, and dusty star-forming galaxies, respectively. 
   In the case of non-merging galaxies, these percentages are 
  30.5$\pm$0.7\%, 28.8$\pm$0.7\%,  and 40.7$\pm$0.8\%, respectively. This indicates that overall, the sample of major mergers contains a similar fraction of quiescent, unobscured, and dusty  galaxies than non-mergers.

\begin{figure}[!htbp]
\begin{center}
\includegraphics[angle=0,scale=0.40]{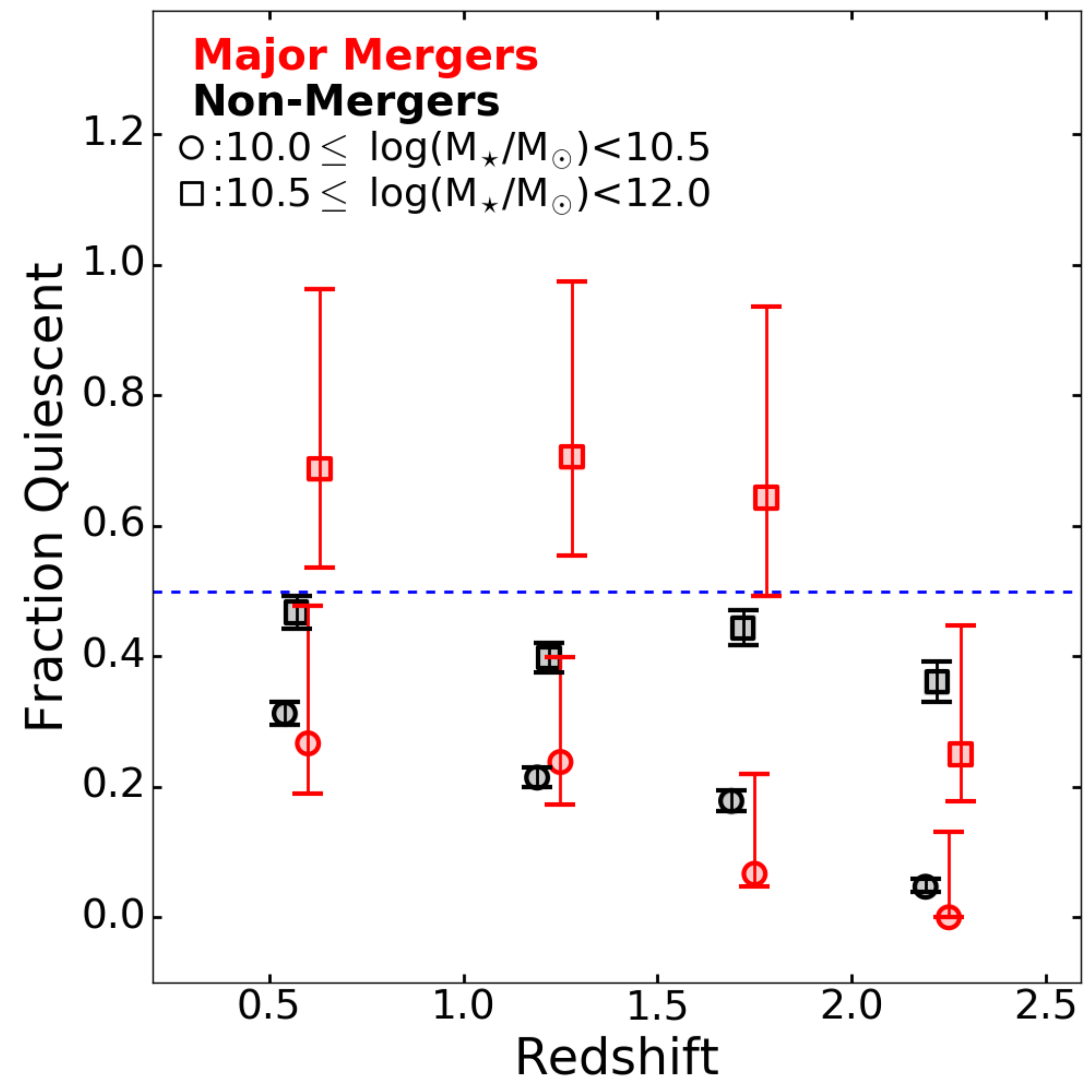}
\caption{Fraction of quiescent galaxies in the high-mass sample of major mergers (red) and in non-merging systems (black) as a function of redshift and in two stellar mass ranges.  Errors are Poisson statistics \citep{gehrels1986}.  The quiescent fraction is lower (there are more star-forming galaxies) at lower masses. The quiescent fraction in the lower mass bin decreases with increasing redshift, and is similar for both merging and non-merging galaxies. In the high mass range, merging galaxies have a higher quiescent fraction than their non-merger counterparts, except in the highest redshift bin.
   \label{fig_fracq}}
\end{center}
\end{figure}

  Figure \ref{fig_fracq}  shows the fraction of quiescent galaxies in the high-mass  major merger sample as a function of redshift in two bins of stellar mass. These fractions are compared with the fractions of quiescent galaxies in non-merging galaxies. 
 The fraction of quiescent galaxies with 10$\le\log(\rm M_{\star}/\rm M_{\odot})\le$10.5 in merging systems is similar to the quiescent fraction in non-merging galaxies, and it is found to increase with decreasing redshift.   Similarly to non-merging galaxies \citep[e.g.][]{martis2016}, at $\log(\rm M_{\star}/\rm M_{\odot})<$10.5 star-forming galaxies dominate the population of galaxies in mergers. 
 Figure \ref{fig_fracq}  also suggests  that at $z<2$ and masses $\log(\rm M_{\star}/\rm M_{\odot})\ge 10.5$, there is a higher fraction of quiescent merging  galaxies 
   compared with the fraction of quiescent non-merging galaxies, although it is not significant given the large errors at all redshifts. 
   At $z > 2$ the fraction of quiescent galaxies in the two populations
    is similar within the uncertainties and shows a higher fraction of star-forming galaxies.  The fractions in this mass range are dominated by galaxies with masses $10.5< \log(\rm M_{\star}/\rm M_{\odot})<11.0$. 
 The higher fractions of quiescent merging galaxies at high masses compared to non-merging galaxies 
 could be a consequence of the adopted peak-finding code, which is more sensitive to the selection of bulge-dominated galaxies (more prominent at the high-mass end)  and also to the mass ratio cut which tend to select more quiescent galaxies (see Appendix  \ref{comparison_ratios})).
 There is no clear trend in the evolution of quiescent galaxies in both merging and non-merging galaxies as a function of redshift in this high-mass range.

Table \ref{tbl_qsf} shows the percentages of quiescent, unobscured, and dusty star-forming galaxies in the high-mass major merger sample at different redshifts compared   
with the percentages for non-merging galaxies.   
The fraction of quiescent galaxies  in mergers and non-mergers is similar in most redshift bins and decreases with redshift. 
 The fraction of unobscured star-forming galaxies in mergers and non-mergers  increases with redshift. The fractions for mergers and non-mergers at $z<2$ are consistent with the errors.  At $z>2$ this  fraction is higher in non-merging galaxies. 
In the case of dusty star-forming galaxies,  the fraction in mergers is found to increase with increasing redshift, whereas this decreases  in non-merging galaxies. 
This population is similar between merging and non-merging galaxies at $z<2$, but at higher redshifts, dusty star-forming galaxies are more common among merging than non-merging galaxies.

\begin{deluxetable*}{lcccccc}
\tabletypesize{\footnotesize}
\tablecaption{Percentages$^{a}$ of galaxy types  in major mergers and non-merging galaxies  with masses $\mgediez$ in different redshift bins.  \label{tbl_qsf}}
\tablewidth{0pt}
\tablehead{
\colhead{$z$} &  \multicolumn{2}{c}{Quiescent} & \multicolumn{2}{c}{Unobscured SF} &  \multicolumn{2}{c}{Dusty SF}\\
\colhead{} &   \colhead{Merging} & \colhead{Non-merging}  &  \colhead{Merging} & \colhead{Non-merging} & \colhead{Merging} & \colhead{Non-merging}   
}
\startdata
 0.3-1.0        &  48.4$\pm ^{16.0}_{9.5}$ & 37.8$\pm$1.5      &  9.7$\pm ^{9.4}_{2.9}$ & 17.8$\pm$1.0            & 41.9$\pm ^{15.1}_{8.7}$ &   44.4$\pm$1.6 \\
1.0-1.5       &  44.7$\pm ^{13.7}_{8.4}$ &29.6$\pm$1.3    &  23.7$\pm ^{10.8}_{5.5}$ & 26.3$\pm$1.2              & 31.6$\pm ^{12.0}_{6.7}$ & 44.1$\pm$1.6   \\
1.5-2.0        & 34.5$\pm ^{14.7}_{7.8}$ & 29.9$\pm$1.5    &   24.1$\pm ^{13.0}_{6.1}$ & 32.0$\pm$1.5               & 42.4$\pm ^{15.7}_{8.8}$ & 38.1$\pm$1.7   \\
2.0-2.5        &  13.3$\pm ^{10.5}_{3.9}$ & 18.1$\pm$1.4   &  30.0$\pm ^{13.7}_{7.0}$ & 52.1$\pm$2.4              & 56.6$\pm ^{17.3}_{10.7}$ & 29.8$\pm$1.8   \\ \hline
 Total          & 35.9$\pm$5.3 & 30.5$\pm$0.7                      &  21.9$\pm 3.4$ & 28.8$\pm$0.7            & 42.2$\pm5.7$ & 40.7$\pm$0.8  \\
 \enddata
\tablenotetext{a}{Errors are calculated using Poisson statistics \citep{gehrels1986}.}
 \end{deluxetable*}

\subsection{Merger Types}\label{sec_mtype}

Since we use the UVJ colors to separate galaxies into star-forming  and quiescent, we can separate mergers into wet (both galaxies are star-forming, i.e. dusty or unobscured),   mixed (one quiescent and one star-forming), and 
dry (both are quiescent).  We find that  53.1$\pm 6.4$\%, 21.9$\pm^{4.9}_{3.4}$\%, and 25.0$\pm^{5.2}_{3.7}$\% of the major mergers correspond to wet, mixed, and dry mergers, respectively.  
 For galaxies in the primary sample, the percentages of these types of mergers are 72.3$\pm$5.3\%, 15.2$\pm^{2.8}_{2.1}$\%, and 12.5$\pm^{2.6}_{1.8}$\%, respectively.

\begin{figure*}[!htbp]
\begin{center}
\includegraphics[angle=0,scale=0.60]{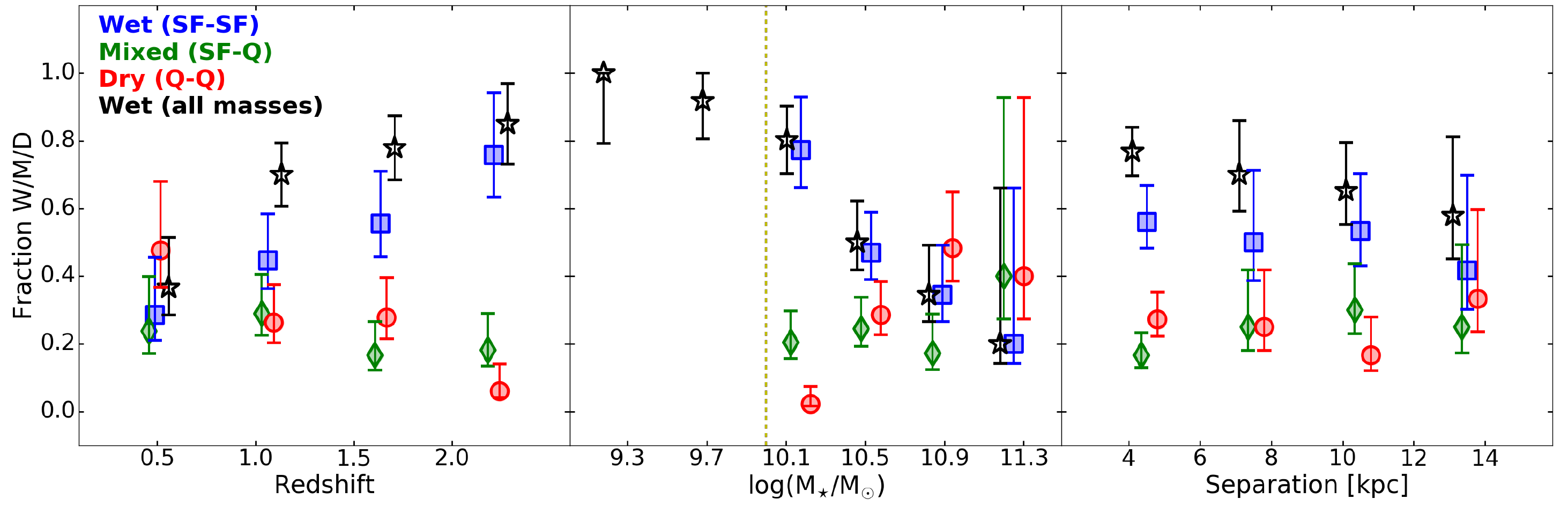}
\caption{Fractions of  wet (blue squares), mixed (green diamonds), and dry (red circles) major mergers as a function of redshift, stellar mass, and  projected separation between the galaxy nuclei. 
The fractions are for the high-mass major merger sample  (both galaxies have $\mgediez$). We include the fraction of wet mergers of the primary sample of mergers (black stars).  The dashed line in the fraction as a function of stellar mass  indicates the separation at $\log(\rm M_{\star}/\rm M_{\odot})=10$.  
 Error bars represent Poisson uncertainties \citep{gehrels1986}.  \label{fig_frac_wmd}}
\end{center}
 \end{figure*}

\begin{figure*}[!htbp]
\begin{center}
\includegraphics[angle=0,scale=0.80]{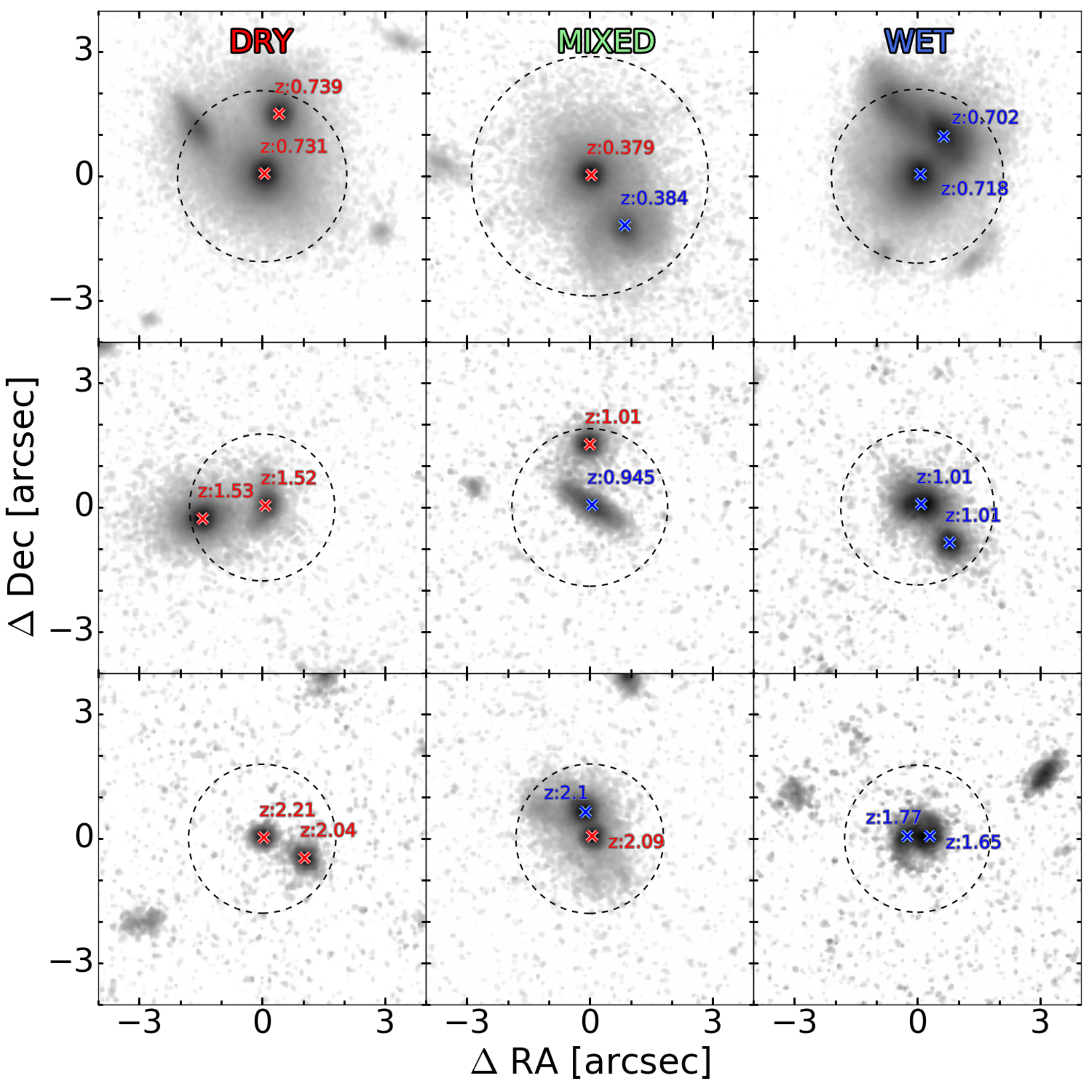}
\caption{Example of merging galaxies selected using the peak-finding algorithm and separated by merger types.
Left, middle and right panels show examples of dry, mixed, and wet  mergers, respectively. The top panels show mergers at $0.3<z\le0.9$, the middle at $0.9<z\le1.7$, and the bottom $1.7<z<2.5$.  Dashed circles shows the area within 15 kpc, in which we search for  mergers. The crosses indicate the position of the identified galaxy nuclei  in the mergers and their colors indicate if the galaxy is quiescent (red) or star-forming (dusty or unobscured, blue). The redshifts of the galaxies are indicated.   \label{fig_merg_wmd}}
\end{center}
\end{figure*}

 Figure \ref{fig_frac_wmd} shows the fraction of the different types of mergers  as a function of redshift, stellar mass, and projected separation between the galaxy nuclei.
 We also analyze the fraction of wet mergers of the  primary sample (black stars).
Wet mergers are dominant at higher redshifts (increment of 48\% from $z \sim 0.5$ to $z=2.5$) 
and lower masses (increment of 57\% from the lowest to the highest mass bin).
The fraction of wet mergers is found to be increasing marginally with decreasing projected separation between the galaxy nuclei (from 0.42$\pm ^{0.28}_{0.11}$ to 0.56$\pm^{0.11}_{0.08}$ at separations of 12-15 kpc and 3-6 kpc, respectively).
The same trends in the fraction of wet mergers are seen in the primary sample. 
The fractions of wet mergers for this sample  as a function of the different parameters are slightly higher than in the high-mass sample, although not significant given the error bars.  For wet mergers in the primary sample, the fraction as a function of stellar mass increases with decreasing stellar mass, with all mergers being wet at masses $\log(\rm M_{\star}/M_{\odot} )\sim 9.3$, suggesting that the population of low-mass galaxies in mergers is dominated by star-forming galaxies.
 In the case of mixed and dry mergers, their distributions are similar in most cases. 
They show an increment with decreasing redshift (for dry mergers it increases 42\% from $z>2$ to $z\sim 0.5$) 
and  a constant fraction with nuclei separation ($\sim$0.25). 
The fraction of dry mergers increases with mass (38\% of increment from the lowest to the highest mass bin). 
Figure \ref{fig_merg_wmd} shows examples of wet, mixed, and dry mergers.

\subsection{Star-Formation Activity in Major Mergers}

We compare the star formation activity in merging and non-merging galaxies to analyze if there is an enhancement in the level of star formation activity in mergers. 
 The star formation rate in non-blended galaxies in 3D-HST comes from either the modeling of their SEDs or from the analysis combining the UV+IR emission. 
 In originally blended galaxies, the star formation rate comes primarily from the modeling of their de-blended SEDs, except when the blended system has MIPS 24 $\mu$m detection. In this case, we assigned a new de-blended IR emission to these galaxies in the following way. If the blended system is mixed (i.e., made of a quiescent and a star-forming galaxy), then the  IR emission was assigned entirely to the star-forming component. If the blended system is wet (i.e., made of two star-forming galaxies), then both components were assigned a fraction of the total derived IR emission based on the ratio of their SFRs as derived from the modeling of the SEDs.

\begin{figure*}[!htbp]
\begin{center}
\includegraphics[angle=0,scale=0.70]{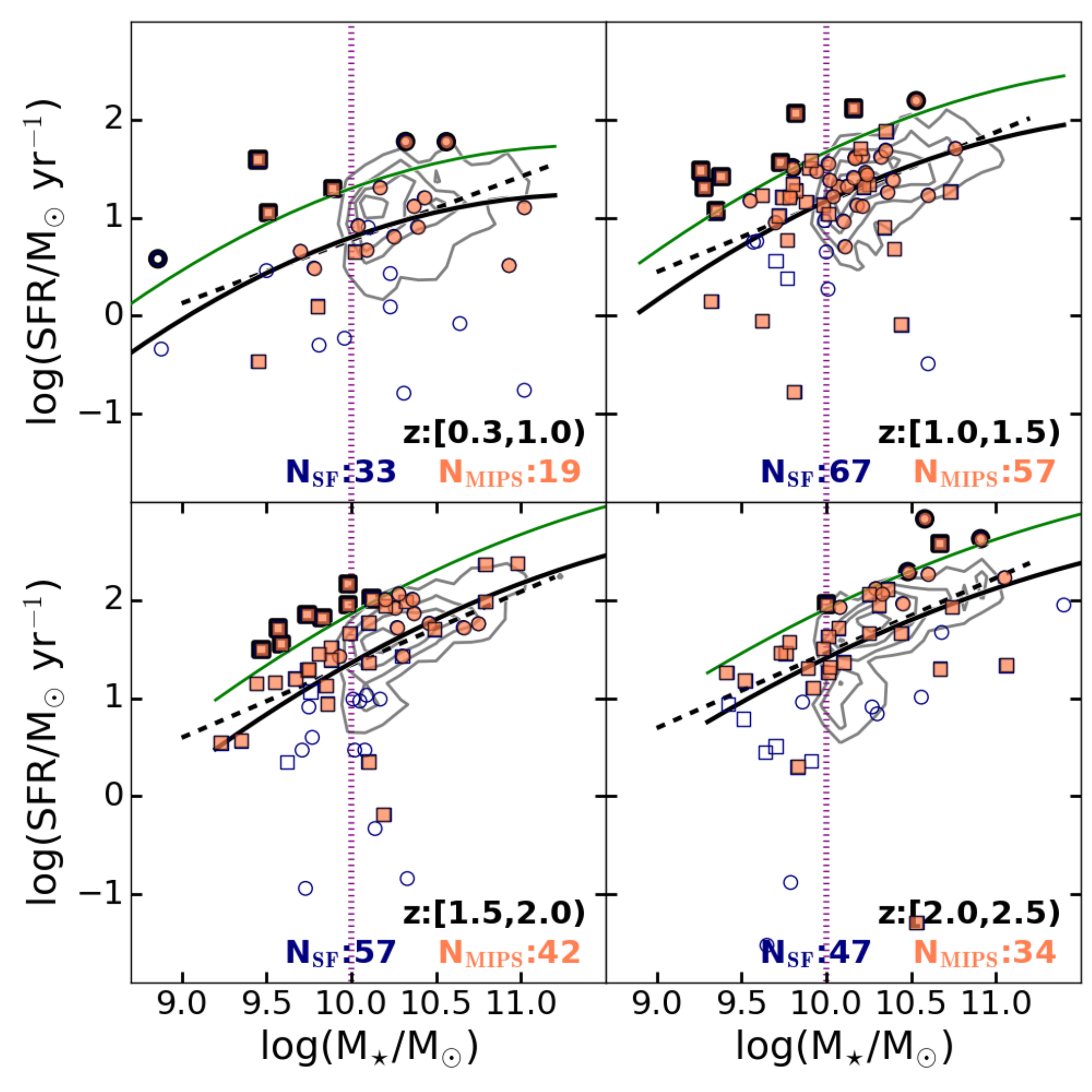}
\caption{Comparison of the star formation rate as a function of stellar mass for star-forming non-merging  and merging galaxies  in different redshift bins. 
 The distribution of non-merging star-forming galaxies  with $\mgediez$ are shown in contours (20, 40, 60, and 80\% of the maximum in the distribution)  while galaxies in mergers are shown with the symbols. Squares are galaxies that were originally blended in 3D-HST and circles are those non-blended. Galaxies with MIPS detection are shown with the filled symbols in orange.   
The dashed and solid curves
 correspond to the star formation  main sequence fits presented in \citet{speagle2014} and \citet{whitaker2014}, respectively.   The green solid curves are 0.5 dex above the main sequence of \citet{whitaker2014}. The number of star-forming galaxies in mergers and those with  MIPS detection is indicated  in each bin. 
Merging galaxies that are $\ge$0.5 dex above the main sequence fit of \citet{whitaker2014} are marked with the thick black symbols.
 \label{fig_mseq1}}
\end{center}
\end{figure*}

\begin{figure*}[!htbp]
\begin{center}
\includegraphics[angle=0,scale=0.70]{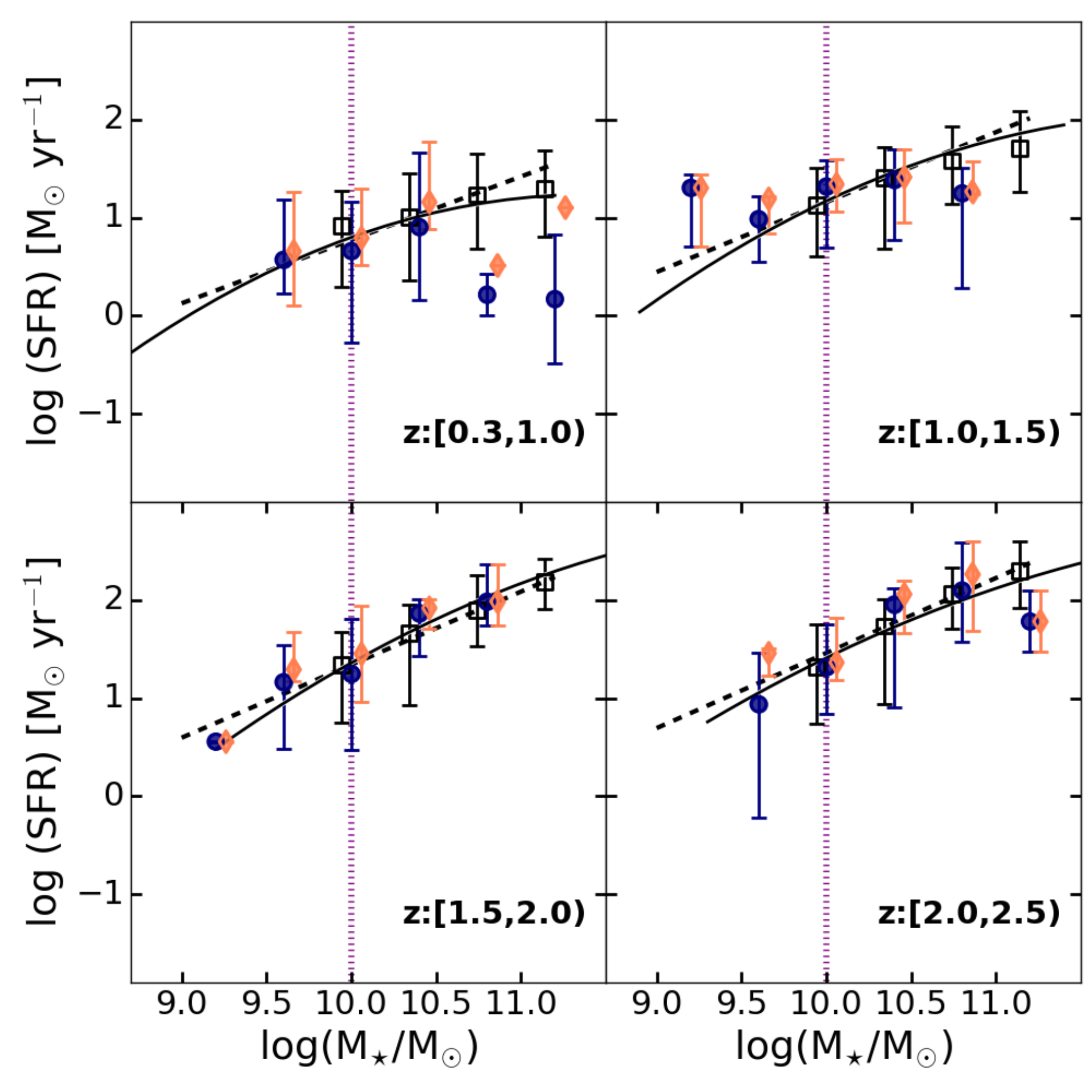}
\caption{
   Median of the  star formation rate as a function of stellar mass for merging and non-merging star-forming galaxies.  Lower and upper error bars correspond to the 15 and 85 percentiles of the distribution. 
 Non-mergers are in open black squares,  merging galaxies are shown with the blue circles, and merging galaxies with MIPS detection are shown with the orange diamonds.  The dashed and solid curves
 correspond to the star formation  main sequence fits presented in \citet{speagle2014} and \citet{whitaker2014}, respectively. 
 \label{fig_mseq2}}
\end{center}
\end{figure*}

Figure \ref{fig_mseq1} shows the position of individual merging galaxies in the star formation rate vs stellar mass diagram including those of the primary sample of major mergers.
We compare the position of the star-forming galaxies  with the main sequence fits obtained from \citet{speagle2014} and \citet{whitaker2014} in different redshift bins.
We also indicate  those galaxies that have MIPS detection because the main sequence fit of \citet{whitaker2014} was obtained from MIPS detected star-forming galaxies in the 3D-HST catalog. 
Figure \ref{fig_mseq2}  compares the median SFRs of merging and non-merging galaxies 
in different stellar mass bins.

We performed a two-dimensional Kolmogorov Smirnov (K-S) test between the merging and non-merging samples  in the SFR vs M$_{\star}$ diagram, to quantify whether the two samples are consistent with coming from  the same parent population.  The resultant  P-values (probability of the plausibility of the null hypothesis)
are tabulated in Table \ref{tbl_kstest}. 
For galaxies with $\mgediez$ (in the high-mass sample), we found that the major merger and the non-merger populations 
 are indistinguishable at any redshift.  
   In the case of galaxies in major mergers with masses $\mltdiez$,
 we find that at $1.0<z<2.0$ the population of major mergers is significantly different to non-mergers in the same mass range\footnote{And in the same redshift range and with $m_{\rm AB}\le24.5$}  (P-values$<$0.003). 
The finding that the population of merging galaxies with $\log(\rm M_{\star}/\rm M_{\odot})$ $< 10$ is different from the population of non-mergers, can be also visualized in  Figure  \ref{fig_mseq1}. This figure shows that low-mass merging galaxies have enhanced star formation compared to star-forming galaxies on the main sequence. In Figure \ref{fig_mseq1},  we  highlight star-bursting merging galaxies, which are those that have SFR larger by more than 0.5 dex with respect to the main sequence fit of 
\citet{whitaker2014}.

In the high-mass sample of mergers, ten galaxies  ($\sim$12\% of the major mergers)  are star-bursting galaxies. 
All of these starbursts are in wet mergers, are dusty star-forming galaxies, and have MIPS detections.  They are in mergers with projected nuclei separation between 4.0 and 11.5 kpc. Four of these galaxies were originally blended in 3D-HST.  
Eight of these star-bursting galaxies have a dusty star-forming galaxy as a companion. 
In the case of major mergers with $\mltdiez$, the fraction of star-bursting galaxies  increases to $\sim$20\%, suggesting
a higher enhancement in the star formation activity in low-mass merging galaxies. 

\begin{figure}[!htbp]
\begin{center}
\includegraphics[angle=0,scale=0.40]{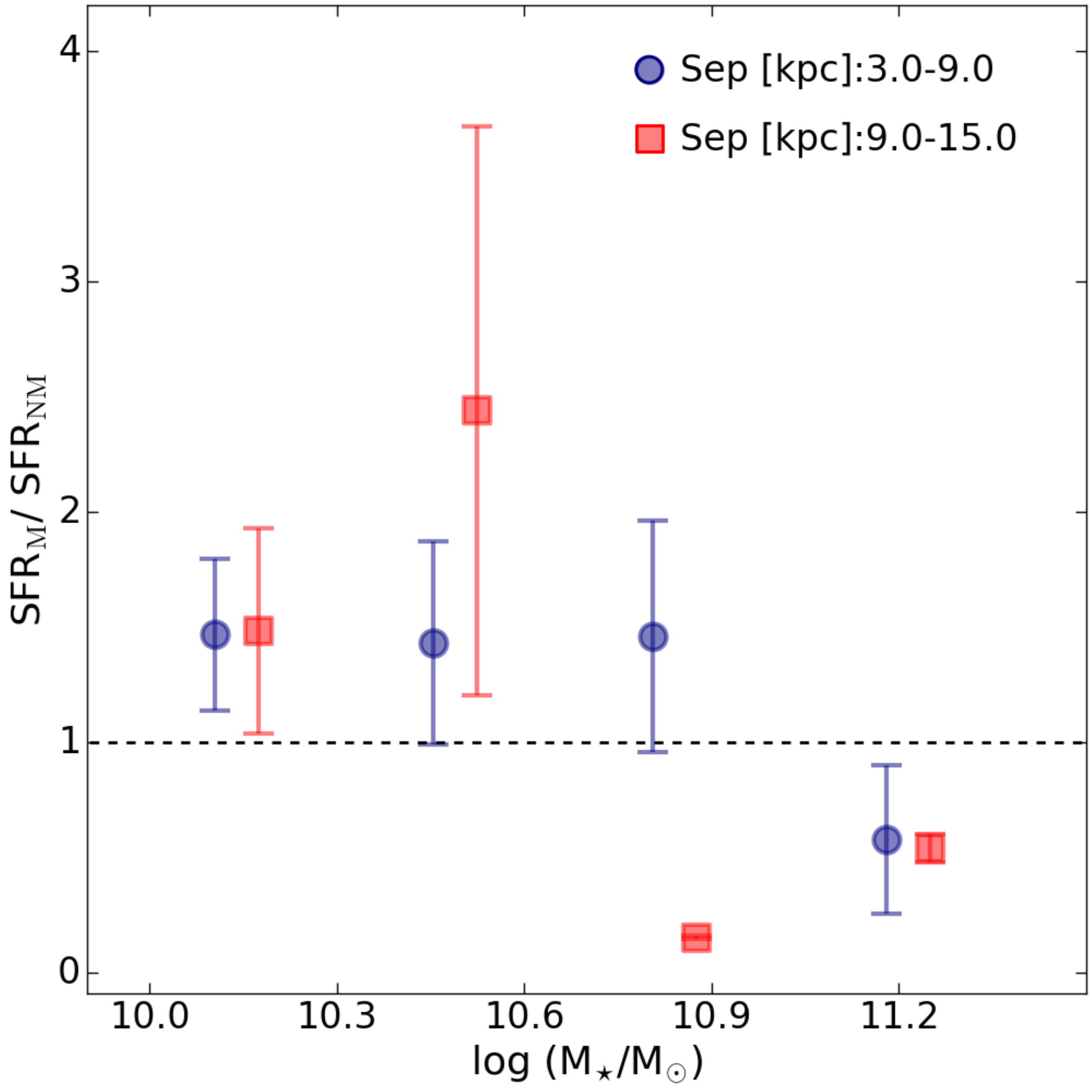}
\caption{ Ratio of the means of the star-formation rate of star-forming galaxies in major mergers (SFR$_{\rm M}$)  with respect to non-merging galaxies (SFR$_{\rm NM}$) as  a function of stellar mass for mergers at nuclei separation 3-9 kpc (circles) and 9-15 kpc (squares).      
 \label{fig_ratio_sfr_merg}}
\end{center}
\end{figure}

To analyze if there is an increment in the star formation rate in mergers with decreasing projected separation, we measure the ratio of the means of the SFR in star-forming galaxies in major mergers and non-mergers as a function of stellar mass (Fig. \ref{fig_ratio_sfr_merg}). We compare these ratios at nuclei separations of 3-9 kpc and 9-15 kpc. 
 Although we do not find  an increment in the star formation activity with nuclei separation, we see an increment of this ratio  of the means in galaxies with lower masses, supporting the idea  that lower-mass merging galaxies might suffer a higher impact in the star formation activity.

\begin{deluxetable}{ccc }[!htbp]
\tabletypesize{\scriptsize}
\tablecaption{ P-values$^{a}$ from Kolmogorov-Smirnov Test  for galaxies in the SFR vs M$_{\star}$ diagram. \label{tbl_kstest} }
\tablewidth{0pt}
\tablehead { $z$ &    High-mass$^{b}$ &  Low-mass$^{c}$}  \startdata
0.3-1.0 & 0.313 &        0.006                  \\
1.0-1.5 & 0.201 & 9.991$\times 10^{-8}$  \\
1.5-2.0 & 0.340 & 1.750$\times 10^{-5}$   \\
2.0-2.5 & 0.050 & 0.190                           \\
\enddata
\tablenotetext{a}{ In the cases where P$<$0.003 the null hypothesis that the two samples are drawn from the same parent population is excluded at more than 3$\sigma$.}
\tablenotetext{b}{Galaxies in major mergers with masses  $\mgediez$.}
\tablenotetext{c}{Galaxies in major mergers with masses $\mltdiez$.}
\end{deluxetable}


\subsection{Potential systematics in the selection of the merger sample}

 We select major mergers using a peak-finding algorithm that select galaxy nuclei in an image. We impose restrictions such as differences in redshift and  separation between the identified galaxy nuclei in order to select mergers. In addition, we use a 
 flux and a mass ratio cut $\ge$ 1:4 to select major mergers.  
 However, as indicated in \citet{man2016}, selecting galaxy mergers based on stellar mass or flux ratio can give different results on measured properties of mergers. 
 We repeat the selection of mergers applying different combinations of flux and mass ratio cuts. 
 In Figures \ref{fracq_z_cratios}, \ref{frac_wet_cratio},  and \ref{mseq_cratio} in Appendix \ref{comparison_ratios},  we repeat the measurement of the fraction of quiescent galaxies as a function of redshift (Fig. \ref{fig_fracq}), the fraction of wet mergers as a function of redshift, stellar mass, and separation (Fig. \ref{fig_frac_wmd}), and the median of the SFR of merging galaxies in the SFR vs M$_{\star}$ plot (Fig. \ref{fig_mseq2}) using the different samples of mergers. 
We find no significant difference between the results using the different combinations of ratio cuts. 
 The sample of \citet{man2016}, contains merging galaxies at separations 10-30 kpc and our sample of mergers is close to coalescence, with galaxy nuclei separations $<$15 kpc. 
 It is possible that \citet{man2016} find different results when using  flux and  mass ratio in their sample selection due to the larger separation of the merging galaxies compared to ours. 
 
 \citet{cibinel2015} indicate that samples of merging galaxies identified from H160 images alone are not ideal because they can have a contamination from clumpy galaxies up to 50\%.  
 They find that using stellar mass maps produces a cleaner selection of mergers with a contamination in the number of clumpy galaxies $<$20\%.
 \citet{wuyts2012}, also using stellar mass maps, find that the contribution of the mass of clumps to the total stellar mass in galaxies is less than 7\%. 
 It is difficult to quantify the possible contamination of clumpy galaxies in our sample of major mergers, since our sample consists of merging galaxies close to coalescence, thus making it potentially more difficult to distinguish between a star-forming galaxy and a massive clump.
 However, with the mass ratio cut used to select our sample of mergers, we avoid the selection of systems with typical  masses of clumps, since we select mergers in which the lowest mass member 
has at least  25\% of the mass of the main galaxy\footnote{For instance, if we select a galaxy pair in which one galaxy has a mass of 10$^{10}$M$_{\odot}$,  a star-forming clump should have a mass $<$7$\times 10^{8}$M$_{\odot}$, thus a mass ratio of 0.07 which is $<$0.25 mass ratio cut we use to select mergers.}. In addition, most of the analysis shown in this paper is based on merging galaxies in which both components have masses $\log(\rm M_{\star}/M_{\odot})>10$, higher than what is expected for star-forming clumps   (masses $8<\log(\rm M_{\star}/M_{\odot})<10$ 
 
 
 Since we are using only the H160-band  images  to select mergers over the redshift range $0.3<z<2.5$, we are selecting merging galaxies from a wide range of rest-frame wavelengths (from rest-frame $\sim$1.18$\mu$m at $z=0.3$, to rest-frame $\sim$0.44 $\mu$m at $z=2.5$). 
We investigate if the use of a single band selection makes an impact in the trends such as those found in  Figures \ref{fig_fracq} and \ref{fig_frac_wmd}. 
 We apply the peak-finding algorithm to I814-band stamps of galaxies in the 3D-HST catalog at $z<1$, because we want to sample  rest-frame wavelengths longer than the $4000\AA$ break. 
 The rest-frame of the I814 images at $z<1$ corresponds to the rest-frame of the H160-band at $z>1.5$. 
 In Appendix \ref{comparison_bands}, we present the comparison in the fraction  of quiescent galaxies as a function of redshift (Fig. \ref{fracq_z_comp_bands}), the fraction of wet mergers as a function of redshift (Fig. \ref{frac_wet_comp_bands}, left), and the mean in the SFR as a function of M$_{\star}$ diagram (Fig. \ref{frac_wet_comp_bands}, right).  
We find no significant differences in the trends reported in this paper when 
 using the I814 and the H160 band images for the selection of mergers.  We therefore conclude that our results are robust.

\vspace{1cm}
\section{Discussion} \label{sec_discussion}

\subsection{Evolution in the Fraction of Wet, Mixed, and Dry Mergers}\label{sec_disc_fwmd}

In Figure \ref{fig_frac_wmd}, we compare the fraction of wet, mixed, and dry mergers as a function of redshift, stellar mass, and projected separation for the sample of high-mass major mergers (in which both merging galaxies have masses $>$10).  
We find that the majority of the major mergers are wet (53\%) and this fraction is higher (72\%) for the  the primary sample of mergers, which include galaxies with $\log(\rm M_{\star}/M_{\odot})<10$. 
We found that the fraction of wet major mergers increases with redshift.  This increase is due to the higher fractions of star-forming galaxies at high-$z$.  

While wet mergers dominate merging events at $z>1$, the relative importance of dry mergers increases over time, 
 in agreement with previous results\footnote{This trend was not observed by \citet{lackner2014} although we used the same method to select mergers (see also \S \ref{comparison_others}). This is because they could not resolve the individual colors of merging galaxies. They could separate mergers only  into wet and dry (not mixed) based on the color of the merging system.} \citep{khochfar2003, lin2008}. This is driven by the increase in the fraction of  early-type galaxies at later cosmic times. 
 We also found an increase of
the dry merger fraction with increasing 
 stellar masses,  in agreement with the increase of the fraction of 
  quiescent galaxies at the high-mass end.  
  On the contrary, wet mergers are dominant at low masses ($\log(\rm M_{\star}/\rm M_{\odot})< 10.5$), with all merging events being wet  at masses $\log(\rm M_{\star}/\rm M_{\odot})<9.3$.  
  A possible explanation for 
   the lack of dry mergers at  low stellar masses could be that  most low-mass galaxies are star-forming or that 
   low-mass quiescent galaxies are too small 
   to be selected using the peak-finding algorithm.
   Since the code selects a minimum number of pixels above 5$\sigma$ from the smoothed image (Section \S \ref{sec_detection}),  it is possible that  small galaxies do not survive the high-pass filtering and therefore are not selected.  

The fraction of mixed mergers  as a function of redshift, stellar mass, and projected nuclei separation follows a similar trend to the fraction found in dry mergers, i.e., increasing with decreasing redshift and increasing stellar masses, 
 and its evolution is roughly flat with separation. 
  Studying a sample of close pairs in the DEEP2 Galaxy redshift Survey at $0.75<z<1.2$, 
 \citet{lin2010} 
found that 
mixed and dry mergers reside in similar environments. These environments are denser than the environments where wet mergers reside. 
 Although the study of the environments of our merging systems is beyond the scope of this paper,  the similar environments of
 mixed and dry mergers could be the reason why their fractions  follow similar trends.  

According to the semi-analytical models of \citet{khochfar2003} and the observations of \citet{lin2008},   dry mergers are the dominant mechanism responsible for the assembly of the most massive present-day elliptical galaxies, while wet and mixed mergers will produce red galaxies of intermediate mass. 
This is in agreement with the finding that wet mergers tend to have lower stellar masses, while 
dry mergers tend to be more common among the more massive galaxies in our sample. 
Mixed mergers might have a combination of a low-mass star-forming galaxy with a massive quiescent galaxy. 

We found marginal evidence for an increase of the incidence of wet mergers with decreasing projected galaxy separation, while the fraction of dry and mixed mergers is found to remain flat as a function of separation.  
The small increase in the fraction of wet mergers with decreasing  galaxy separation could be potentially caused by the mis-identification of an HII region or a star-forming clump within a star-forming galaxy. 
 However, in the peak-finding algorithm, we avoid the selection of edge on disk galaxies and remove peaks that are too faint compared to the brightest peaks  to avoid star-forming clumps.  In addition,  we avoid the potential selection of clumpy galaxies by imposing the restriction that both merging galaxies have $\log(\rm M_{\star}/M_{\odot})>10$, since typical clumps masses are lower than this limit \citep{guo2012}. 
 Therefore, contamination by star-forming clumps should not be the main reason for the increment. 
Another reason for the tentative increase in the fraction of wet mergers with decreasing galaxy separation 
could be that as merging galaxies approach, there is an increment in the level of activity of star formation.
Hydrodynamical simulations of merging galaxies  \citep[e.g.,][]{barnes1996, dimatteo2007, cox2008}   have shown that torques exerted on the gas in the galaxies as a product of merger events result in a loss of angular momentum in the gas, which then falls towards the galaxy nucleus and leads to significant enhancement of star formation.

\subsection{Lack of Starburst Galaxies at this Merger Stage}\label{disc_starbursts}

Galaxies that lie well above the main sequence  of star-forming galaxies (4 or 5$\times$  above) are often interpreted as starbursts triggered by mergers  \citep{rodighiero2011, silverman2015}. This is expected to occur  especially  in 
major mergers since they apparently  trigger the strongest star formation rate enhancements  \citep{ellison2008, woods2010}.  
On the contrary, galaxies  that lie  on the main sequence are thought to  form stars in a quasi-steady process \citep{scoville2014, scoville2016, bethermin2015} rather than through merger events.
Surprisingly, we find that most of the merging galaxies selected in our work lie on the main sequence of star-forming galaxies rather than being star-bursting galaxies. 
While this result is, at face value, in disagreement with the aforementioned interpretation that starbursts are triggered by major mergers, it can be reconciled using the latest results from simulations of galaxy formation, recent ALMA observations, and some investigations of local mergers.

First, \citet{fensch2017}, using  hydrodynamical numerical simulations, compared  the star-formation activity triggered by high- and low-$z$ mergers, assuming they have gas fractions typical of $z=2$ (f$_{\rm gas}$=0.6) and $z=0$ (f$_{\rm gas}$=0.1) galaxies, respectively.  These  simulations showed that the star-formation rate enhancement 
 induced by the high-$z$ mergers and its duration are
  10 times lower than in low-$z$ mergers and the burst of star-formation activity in high-$z$ mergers occurs only at coalescence.

Second, \citet{puech2014} classified 27 $z\sim0.6$ star-forming merging galaxies from IMAGES-CFS  into three different merger stages: 
A pre-fusion phase in which the two progenitor galaxies can still be identified as distinct components; 
a fusion phase during which the merging galaxies are at the coalescence and reach the peak in star formation; 
and a relaxation phase during which the merger remnant reaches a relaxed dynamical state. 
They studied how galaxies in mergers are distributed across the main sequence as a function of their merger stage. They found that galaxies that lie 1$\sigma$ ($>$0.3 dex) above the main sequence were galaxies in the fusion phase, while the main sequence was populated by merging galaxies in all
three merger phases. 
 They claimed that the SFR in each merger phase can be triggered by processes such as internal instabilities, minor mergers, or gas accretion.
  However, in the fusion phase,  the SFR triggered by tidal torques during the merger is most important. It drives a fraction of the gas inward, resulting in a central starburst. 
  
 The nuclei of our merging galaxies have  projected separation between 3-15 kpc, therefore they are still in a pre-coalescence/pre-fusion\footnote{It is possible that they are in a stage after the first pericenter passage. With our current data we cannot distinguish between merging galaxies before or after this stage because we need to make a detailed study of the gas and the metallicity in the merging systems. 
  However, according to the simulations of \citet{fensch2017}, the SFR in high-$z$ mergers does not increase significantly at the first pericenter passage, it is only enhanced at coalescence.  Therefore, it is correct to assume that the mergers are in a pre-coalescence stage, since they show two distinct nuclei.} stage and may be yet to  reach the maximum level of star formation activity, in agreement with the simulations of \citet{fensch2017}  and the observational results of \citet{puech2014}.  However, we found that ten of our pre-fusion     phase mergers   with $\mgediez$  are star-bursting since they have  star-formation rates $>$0.5 dex larger than the star-formation rates of main sequence galaxies.  Therefore, an additional ingredient may need to be taken into account for the increment of the SFR in merging galaxies.  
 
 \citet{scoville2016} estimated the gas mass in $1<z<6$ galaxies in the COSMOS field by using dust continuum observations with ALMA. They suggested that high-$z$ starburst galaxies are the result of their  larger gas masses
 (compared to main-sequence star-forming galaxies at the same redshift) rather than having an  increasing efficiency in converting gas to stars as a result of merging events. 
The ten starbursts with $\mgediez$ found in this work are all dusty star-forming galaxies in wet mergers and eight of them have a dusty star-forming galaxy as a companion. 
Studies have shown that  there is a correlation between the gas and dust mass in galaxies \citep[which also depends on the metallicity and stellar mass of the galaxy, e.g.][]{magdis2011}.
If the dust content is a proxy of the gas mass in galaxies,  it is possible that our sample of star-bursting merging galaxies has higher gas fractions (due to the higher dust content in the merging system)  than the merging galaxies that lie on the main sequence.
With higher gas fractions, galaxies in mergers  could deviate above the main sequence even at early stages.  
Although our sample of starbursts is small, our results suggest that  the enhancement  of star formation activity depends on both the merger phase as indicated by \citet{fensch2017} and  \citet{puech2014}  and the gas content in the merging system as indicated by \citet{scoville2016}. 
Galaxy mergers  containing higher gas (dust) fractions will reach the starburst phase earlier than at coalescence, although the level of star formation enhancement and its duration could arguably be reduced compared to local mergers. 
 It is possible that the  majority of the observed starbursts at high-$z$ are produced by (non-merging) galaxies with high gas fractions as indicated  by \citet{scoville2016} and only a small fraction of starbursts are triggered by major mergers.  
 The starburst phase produced by mergers might be of short duration (as indicated by simulations) and may also depend on the gas fraction of the merging galaxies, making it   more difficult to identify starbursts produced by a merging  process.


\subsection{Impact on the star formation activity in member galaxies}
We note from Figure  \ref{fig_mseq1}, that at masses $\mltdiez$ there is 
a significantly larger incidence of starbursts  than at larger stellar masses (20\% and 12\% of the major merging galaxies are starbursts in the low- and high-mass range, respectively). 
 It might indicate at first glance, that the star formation activity in
merging galaxies with the lower mass are more affected by the interaction than their higher mass companions. 
To know which of the merging galaxy in our sample are affected the most, we compare the specific star formation rate (sSFR=SFR/M$_{\star}$) of the lower and higher mass member galaxy in wet mergers.
When both merging galaxies have masses $\log(\rm M_{\star}/M_{\odot})<10$, we find that 31\% of the lower mass  merging members have higher sSFR than their higher mass  companions. When at least one galaxy in the merger has a mass $\log(\rm M_{\star}/M_{\odot})>10$, this fraction increases to 37\%. 
These results indicate that  the most massive merging galaxy is more affected by the interaction, in agreement with the findings of 
\citet{davies2015}.  In their work, they investigated the link between dynamical phase and star formation in galaxy pairs at $z<0.3$. 
They found that star formation is enhanced in the higher mass companion in the pair and suppressed in the lower mass galaxy. 
They suggest that the suppression in star formation in the low-mass galaxy is due to gas heating or stripping, while the enhancement in the high-mass galaxy is produced by tidal gas turbulence and shocks. 
In contrast,  we find that the fraction of mergers in which the lower mass member has higher sSFR than its higher mass  companion is  56\% when at least one member has a mass $\log(\rm M_{\star}/M_{\odot})>10.5$, in agreement with \citet{li2008}. 
They 
studied the interaction-induced star formation in a sample of 10$^{5}$ star-forming galaxies at $0.01\le z \le0.3$. 
They measured the enhancement in star formation as a function of projected separation between two galaxies and found that the star formation in low-mass galaxies ($\langle \log (\rm M_{\star}/\rm M_{\odot})\rangle$=9.72) is more strongly enhanced than in the most massive galaxies ($\langle \log (\rm M_{\star}/\rm M_{\odot})\rangle$=10.6).
Our findings suggest that as  the mass of both merging galaxies increases,  the star formation activity in the less massive member is more affected by the interaction.

\subsection{Comparison with previous works} \label{comparison_others}

In our work, we adopt the method presented in \citet{lackner2014} to find merging galaxies. \citet{lackner2014} select galaxy pairs  with projected separations 2.2-8.0 kpc 
from a mass-complete sample (merging systems with total mass $\log (\rm M_{\star}/M_{\odot}) >10.6$)  in the redshift range $0.25<z<1.0$.  They apply the  peak-finding algorithm  to ACS I-band images for galaxies in the COSMOS field. In Table \ref{tbl_comp_pfinder},  we compare the fraction of merging galaxies found in \citet{lackner2014} and in our work using the H160 images in the CANDELS fields (we compare with the primary,  the high-mass sample, and with merging galaxies in which both components have masses $\log(\rm M_{\star}/M_{\odot})>10.6$). 
Since \citet{lackner2014} did not resolve the properties of  individual galaxies in pairs, they correct the fraction of mergers for chance superpositions, incompleteness, and contamination from minor mergers and clumpy galaxies using simulations. 
In our approach, instead, we model the de-blended SEDs of the individual merging galaxies, deriving new photometric redshifts and stellar population properties. Moreover, our high-mass sample is complete at $\log(\rm M_{\star}/M_{\odot})>10$ for the individual galaxies in the merging systems, whereas the sample of \citet{lackner2014} is complete for merging systems with total mass $\log(\rm M_{\star}/M_{\odot})>10.6$.
 In addition, the sample of mergers in \citet{lackner2014} has merging galaxies with nuclei separations from 2.2 to 8 kpc and in our work  from 3 to 15 kpc. 
As shown in Table \ref{tbl_comp_pfinder}, we find a lower fraction of mergers compared to \citet{lackner2014} in the redshift range in which the samples overlap. 
Given the mentioned differences in the two samples, it is difficult to pin down where these differences arises from.

\begin{deluxetable*}{lccccccccc }[!htbp]
\tabletypesize{\scriptsize}
\tablecaption{Fraction of mergers ($f$ in \%) as a function of redshift obtained using the peak-finding algorithm.\label{tbl_comp_pfinder} }
\tablewidth{0pt}
\tablehead {
\colhead{$z$} & \multicolumn{3}{c} {\citet{lackner2014}} & \multicolumn{3}{c}{High-mass sample}   & \multicolumn{3}{c}{Primary sample}\\ 
\colhead{ } &  \colhead{N$_{\rm gal}$} &  \colhead{N$_{\rm mgr}$ } &  \colhead{$f$ (corrected$^{a}$)  } &  \colhead{N$_{\rm gal} ^{b}$} &  \colhead{N$_{\rm mgr}$ } &  \colhead{$f$}  & \colhead{  N$_{\rm gal}^{c}$ } & \colhead{ N$_{\rm mgr}$   } & \colhead{ $f$   }
} 
\startdata
 $ [0.25,0. 45)^{d}$   & 867  & 15  & 1.7$\pm$0.6 (1.9$\pm$0.5) & 128  & 0 & 0 (0)$^{*}$  & 1039 & 2& 0.2$\pm$0.3 \\
 $[0.45, 0.7)$        &  1644  & 39      &    2.4$\pm$0.4 (4.2$\pm$0.8)          & 597   & 8 &   1.3$\pm$0.7  (1.6$\pm$1.3)$^{*}$ & 3168 & 10& 0.3$\pm$0.1 \\
$  [0.7, 1.0)  $      &  3383  &   82     & 2.4$\pm$0.3   (6.6$\pm$0.8)     & 1066 &  8  & 0.7$\pm$0.4 (1.1$\pm$0.9)$^{*}$   & 4317 & 15& 0.3$\pm$0.1  \\
 $[1.0, 1.5)   $      &  --      &      --    &   --         &  1785 &  19 & 1.1$\pm$0.3  & 4952 & 42& 0.8$\pm$0.1  \\
$[1.5, 2.0) $         &  --      &      --    &    --          &  1374 & 14  & 1.0$\pm$0.4  & 3094 & 35& 1.1$\pm$0.2  \\ 
$ [2.0, 2.5]$         &  --      &     --     &    --          &   895 &  15 &  1.7$\pm$0.6  & 1753 & 26& 1.5$\pm$0.4  \\
\enddata
\tablenotetext{a}{Corrected by chance superpositions, incompleteness, minor mergers, and clumpy galaxies.}
\tablenotetext{b}{Note that the total number of galaxies is higher than the original sample of 5717 galaxies with $\log(\rm M_{\star}/M_{\odot})>10$, since now we include deblended sources. }
\tablenotetext{c}{Galaxies with masses  $\log(\rm M_{\star}/M_{\odot})>8.3$, because the least massive galaxy in the primary sample of mergers has that mass. }
\tablenotetext{d}{In our work, the range is from 0.3$<z<$0.45. }
\tablenotetext{*}{Only galaxies with masses $\log(\rm M_{\star}/M_{\odot})>10.6$.}
\tablenotetext{}{Errors are obtained from propagation of Poisson statistics.} 
\end{deluxetable*}

\begin{figure}[!htbp]
\begin{center}
\includegraphics[angle=0,scale=0.40]{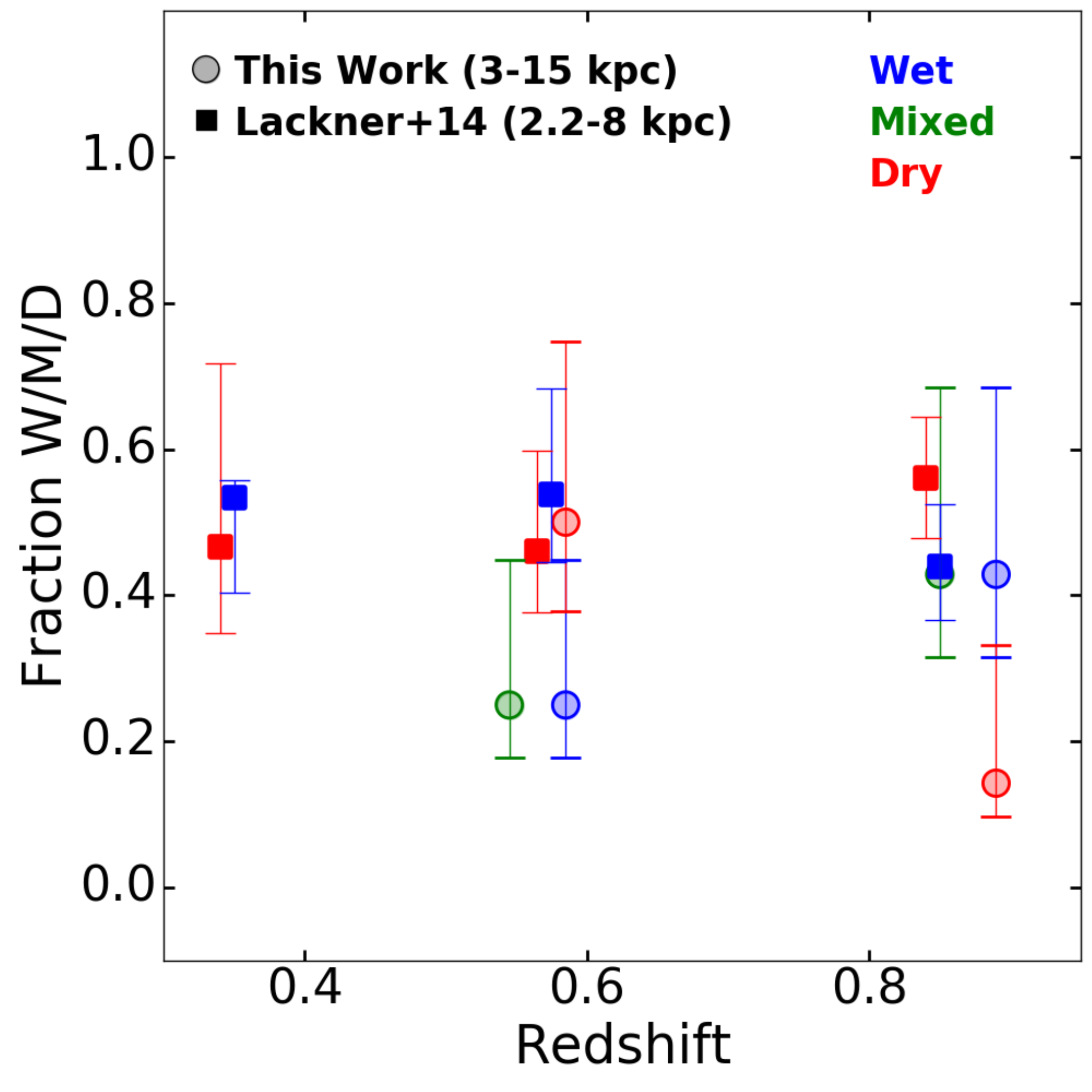}
\caption{ The fraction   of wet and dry mergers as a function of redshift found in this work is shown in circles (high-mass sample) for mergers with galaxy nuclei separations 3-15 kpc  and in \citet{lackner2014} (squares) at separations 2.2-8.0 kpc  in the redshift range where the samples overlap ($0.3<z<1$). Contrary to \citet{lackner2014}, we could resolve the colors of individual galaxies in mergers, thus we could construct a  sample of mixed mergers (included in the plot). 
 \label{frac_wmd_others}}
\end{center}
\end{figure}

\citet{lackner2014} separate their sample into quiescent and star-forming based on the rest-frame near-UV (NUV)$- r^{+}$ and $r^{+}-J$ colors, where the NUV corresponds to the GALEX filter at 0.23 $\mu$m and $r^{+}$ refers to the Subaru $r$-band. They classify mergers as wet/dry  if the total merging system is star-forming/quiescent, respectively.  
In contrast, we could resolve the colors of individual galaxies in mergers and separate mergers into wet, mixed, and dry as presented in  \S \ref{sec_mtype}. 
 Figure \ref{frac_wmd_others} shows the fraction of wet and dry  mergers as a function of redshift found by \citet{lackner2014} compared to  the fraction of wet, mixed, and dry mergers found in our work. 
\citet{lackner2014} find an almost constant fraction of wet and dry merger with redshift.  In contrast,  we find an increment in the fraction of dry mergers with decreasing redshift, in agreement with previous results (see \S \ref{sec_disc_fwmd}).  
We stress however that our results and those from \citet{lackner2014} cannot be directly compared, because our colors are resolved for the merging galaxies, whereas those from \citet{lackner2014} are integrated over the whole merging system (in addition to the differences in stellar mass of the studied samples). Because we define wet/mixed/dry mergers based on resolved colors, our analysis and findings are arguably more robust



\begin{figure*}[!htbp]
\begin{center}
\includegraphics[angle=0,scale=0.63]{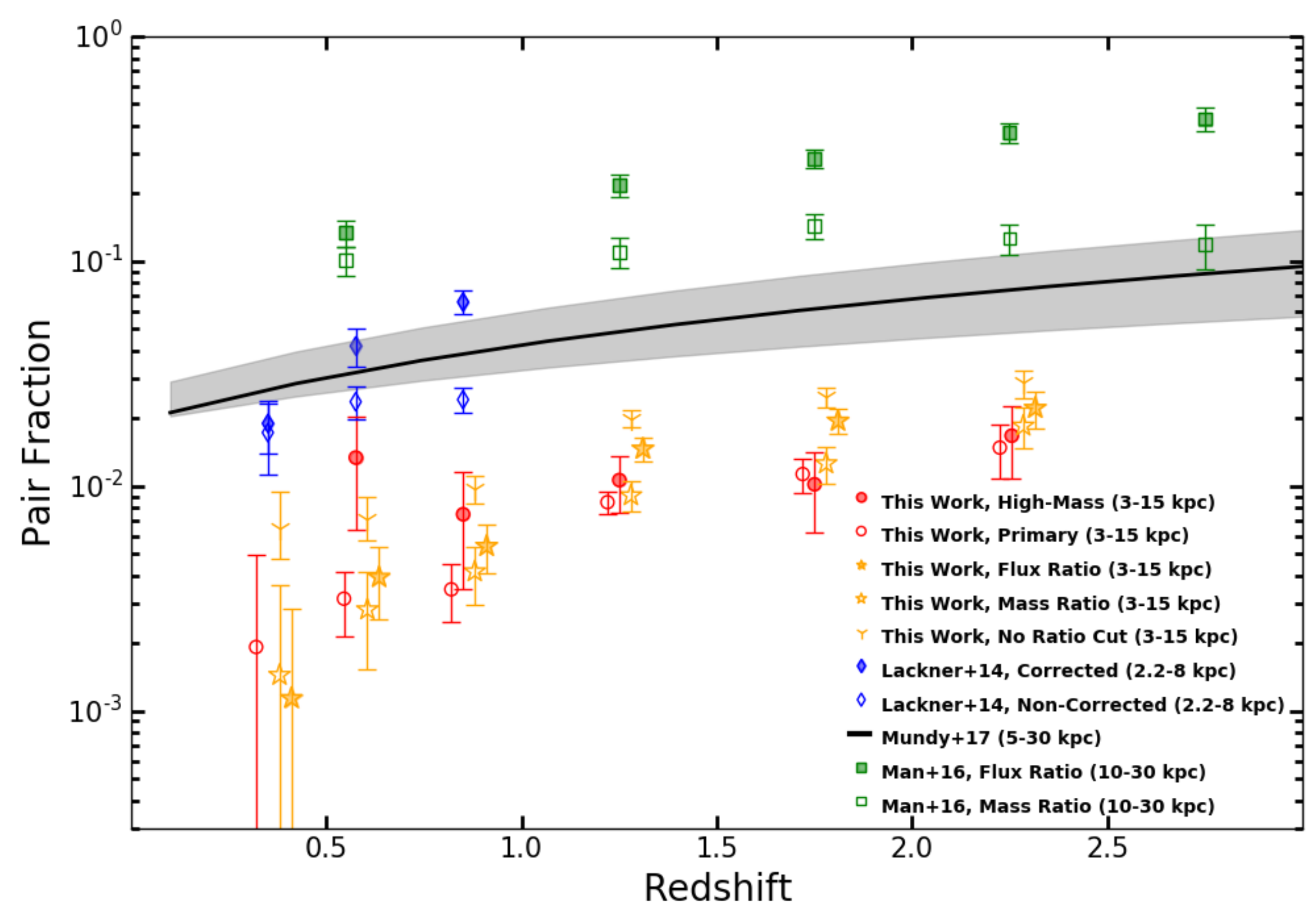}
\caption{ Pair fraction as a function of redshift for samples with different nuclei separations. 
The circles and stars  are for the sample of mergers found in this work, the diamonds  are for \citet{lackner2014}, the squares are for \citet{man2016}, and the solid curve is the fit 
obtained by \citet{mundy2017} for merging galaxies with masses $>10^{10}$M$_{\odot}$. The galaxy nuclei separation of the different samples is indicated.   The pair fractions found in this work are shifted in redshift for better visualization. 
 \label{frac_z_others}}
\end{center}
\end{figure*}

Although the measurement of the merger rate will be addressed in future publications, Figure \ref{frac_z_others} presents a simple comparison with previous works of the pair fraction (value that is necessary to calculate the merger rate) as a function of redshift.  
The pair fraction is defined as the number of pairs (merging systems) divided by the total number of galaxies found in a given redshift bin (in our case $N_{mgr}/N{_{gal}}$, Table \ref{tbl_comp_pfinder}).  
The filled and empty red circles are the fraction of mergers found for the high-mass and the primary sample, respectively.  
We remind the reader that the high-mass sample includes merging systems in which all the individual galaxies have $\log(\rm M_{\star}/M_{\odot})>10.0$, whereas for the primary sample the lowest mass galaxy has $\log(\rm M_{\star}/M_{\odot})>8.3$.
The fraction of mergers obtained when we apply only a flux and a  mass ratio cut, and when we apply no ratio cut (Appendix \ref{comparison_ratios}) is shown with the orange symbols. 
The open and filled diamonds show the pair fraction obtained by \citet{lackner2014} for their corrected and non-corrected values, respectively. 
The open and  filled squares show the pair fraction found by \citet{man2016}\footnote{We plot $N_{\rm major}/N_{\rm massive}$ from  the values  tabulated in Tables 1 and 2 for their sample of mergers obtained from the 3D-HST catalog. }. 
They present a photometrically selected sample of mergers at 0.3$<z<$3 and $\log(\rm M_{\star}/M_{\odot})>10.8$ using  the ULTRAVISTA and the 3D-HST catalogs. They select mergers with projected separations between 10 to 30 kpc using the H160-band flux and stellar mass ratios.  
The solid curve in Figure \ref{frac_z_others} is the best fit in the pair fraction for galaxies found by \citet{mundy2017} ($f_{pair}(z)=(0.019^{+0.007}_{-0.006}) \times (1+z)^{1.16^{+0.042}_{-0.37}}$). 
This fit is for merging galaxies with stellar masses $>10^{10}$M$_{\odot}$ and nuclei separations between 5 to 30 kpc. They obtain the fit from studying a sample of mergers at 0.05$<z<3.5$ from the UKIDSS UDS, VIDEO/CFHT-LS, UltraVISTA/COSMOS and GAMA survey regions. 
Direct comparisons of the pair fractions plotted in Figure \ref{frac_z_others}  is difficult given the different stellar mass ranges probed by the different works and the different separations of the merging galaxies. For example, in addition to the different stellar mass range ($\log(\rm M_{\star}/M_{\odot})>10.0$   vs $\log(\rm M_{\star}/M_{\odot})>10.8$), 
our work and that from \citet{man2016} probe almost complementary regimes in nuclei separation, with d=3-15 kpc and d=10-30 kpc, respectively. However, it appears that the pair fraction decreases drastically as the separation of the nuclei decreases. Finding pairs separated only by a few kpc is observationally challenging, especially at high redshift, and methods like the one developed by \citet{lackner2014}, which we adopted, are necessary to probe this small separation regime. Our work contributes to extend to high redshift ($z\sim$2.5) the investigation of merging galaxies separated by 3-15 kpc and close to coalescence.

\section{Summary and Conclusions} \label{sec_conclusion}

We studied the star formation properties of a sample of merging systems at $0.3<z<2.5$ and having a 
projected separation of 3-15 kpc between the individual nuclei. 
We selected galaxy pairs by applying a peak-finding algorithm to the HST F160W images of sources identified in the CANDELS/3D-HST catalog which covers the COSMOS, AEGIS, GOODS-N, GOODS-S, and UDS fields.  
We find that 28\% of these galaxy pairs are blended (not resolved) in 3D-HST and  obtained the de-blended properties of their members. 
Using  a flux and mass ratio cut $\ge$1:4, and the redshifts and projected separations of the galaxies in pairs, we found 130 major merging systems.
With the combination of the peak-finding code and the de-blended analysis, we increased the sample of major mergers by $\sim$73\% compared to what 
would have been obtained using the 3D-HST catalog only. This increment is substantial for systems with galaxies with stellar masses $\mltdiez$, nuclei projected separations $<$9 kpc, and redshifts $z>1$.
Since the sample of major mergers could be contaminated by star-forming clumps, we select those major mergers in which both merging galaxies have $\log(\rm M_{\star}/M_{\odot})>10$ (64 systems).
Wet, mixed, and dry mergers were classified according to the colors of the individual components in the mergers.   Our findings can be summarized as follow:

\begin{itemize}

\item The majority of the major mergers are wet and their fraction increases with redshift and decreasing stellar mass. 
The fraction of dry mergers increases over time due  to increase of the fraction of early-type galaxies at later cosmic times.

\item We compare the star formation activity in merging and  non-merging galaxies and found  no significant difference between them.   We found that  only $\sim$12\% of the major merging galaxies  with $\mgediez$ are star-bursting, with star-formation rates larger than the main sequence by 0.5 dex or more. These galaxies are dusty star-forming galaxies and are in wet mergers.  
On the contrary, merging galaxies with lower masses showed a higher fraction of star-bursting galaxies (20\%).

 \item  We find that as the mass of the merging galaxies increases,  the star formation activity in the  less massive member in the merger suffers a more dramatic impact than its companion galaxy.

 \item Simulations of high-$z$ mergers suggest that the maximum level of star formation activity in merging systems will occur at coalescence and the intensity and duration of this activity will be lower than in local mergers. 
 Since our sample of mergers have galaxy nuclei separations between 3-15 kpc, they are still  in an pre-coalescence stage and are yet to  reach their maximum level of star formation activity.   

\item  Our results suggest that the enhancement in the star formation activity  in mergers might also depend on the physical properties of the merging galaxies, such as the gas (dust) content and stellar mass. 
  Mergers with higher gas fractions and those with lower stellar masses will increase their star formation activity even before  coalescence. \\
 
 \item Our work contributes to extend to $z\sim$2.5 the investigation of merging galaxies close to coalescence. 
 
 \end{itemize}

We thank the anonymous referee for useful comments which  helped to  improve this paper.

\begin{appendix}

\section{Comparison of the merger properties selected using different mass and flux ratios} \label{comparison_ratios}

We compare the properties of merging galaxies selected using different combinations of  flux and mass ratio cuts. 
We repeat relevant figures such as Figures \ref{fig_fracq}, \ref{fig_frac_wmd}, and \ref{fig_mseq2} applying different restrictions in the selection of mergers:  no restriction in mass and flux ratio (No FR), only mass ratio cut (No FR+MR), only flux ratio cut (FR), and both flux and mass ratio  cut (FR+MR; as adopted for the "primary" sample ). 
In addition to these cuts, we also apply   the restrictions indicated in section \ref{sec_detection} (such as  restrictions in redshift values and  minimum and maximum separation between the nuclei) to select merging galaxies. 

\begin{figure}[!htbp]
\begin{center}
\includegraphics[angle=0,scale=0.40]{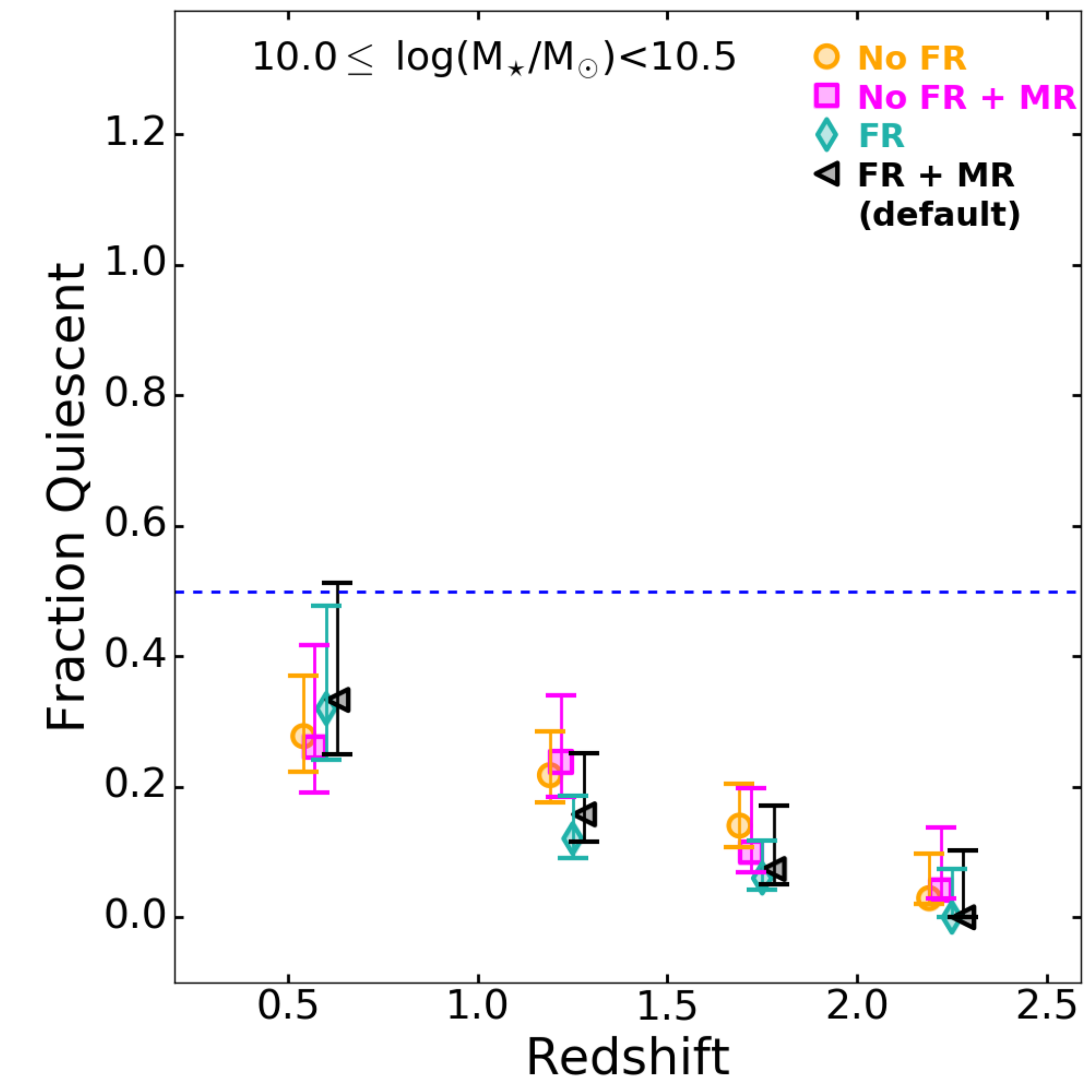}
\includegraphics[angle=0,scale=0.40]{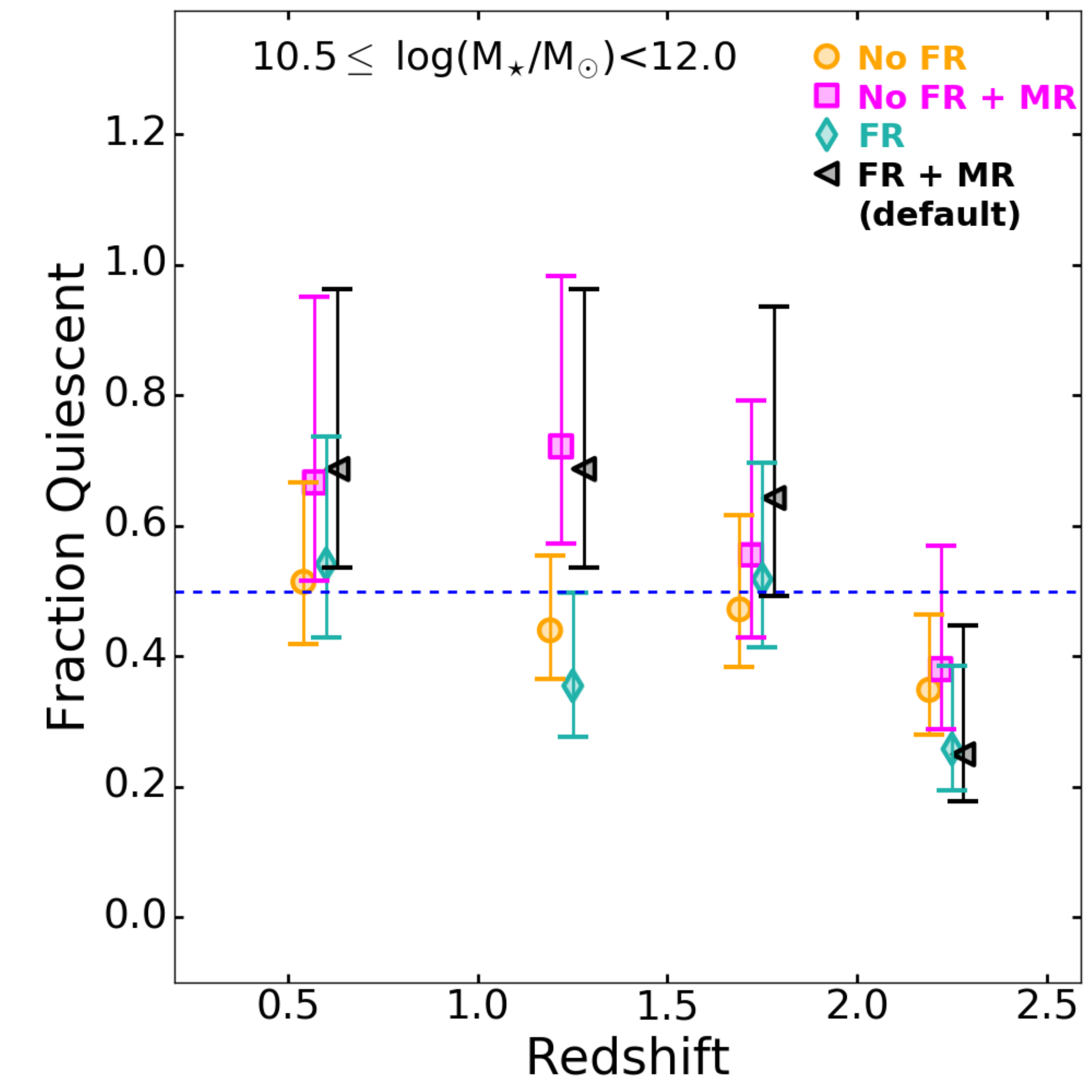}
\caption{ Fraction of quiescent galaxies as a function of redshift for galaxies in mergers with masses $10.0\le \log(\rm M_{\star}/M_{\odot})<10.5$ (left) and $10.5\le \log(\rm M_{\star}/M_{\odot})<12$ (right).
The symbols show the fraction of quiescent galaxies in mergers 
 obtained using no restriction in flux and mass ratio (No FR), only with mass ratio cut (No FR + MR), with only flux ratio cut (FR), and with both flux and mass ratio  cut (FR+MR). 
 We find an increment of the fraction of quiescent galaxies with decreasing redshift for any ratio restriction. In the high-mass range, we do not see a clear trend in the fraction of quiescent galaxies with redshift. Comparing both mass ranges, the fraction of quiescent  galaxies is always lower in the low-mass than in the high-mass range for any ratio cut. 
 \label{fracq_z_cratios}}
\end{center}
\end{figure}

Figure \ref{fracq_z_cratios} shows the fraction of quiescent galaxies as a function of redshift for two mass bins as in Figure \ref{fig_fracq}.
In the mass range $10.0\le \log(\rm M_{\star}/M_{\odot})<10.5$, the maximum difference between the less restrictive cuts (FR, no FR+MR, FR) with respect to the cut that includes both flux and mass ratio (FR+MR) is smaller than the uncertainties in the measurements (differences are 47\% to 97\% of the  values of typical uncertainties).  In this mass range, we see an increment in the fraction of quiescent galaxies with decreasing redshift using any flux or mass ratio cut. 
In the mass range $10.5\le \log(\rm M_{\star}/M_{\odot})<12$, the differences in the fractions are smaller than the typical uncertainties except in the redshift range $1<z<1.5$, where the mass ratio restriction is the major factor that makes the difference. 
In this mass range, we do not see a clear trend in the fraction of quiescent galaxies as a function of redshift  for any mass or flux ratio cut. 
At $z>2$ the fraction of quiescent galaxies is smaller than the fraction of star-forming galaxies. 
In most cases, the fraction of quiescent galaxies tend to be higher when the mass ratio cut is applied. 
Comparing both mass ranges, the fraction of quiescent galaxies is  always lower in the low-mass  than in the high-mass range, independent of the adopted ratio restriction. 

\begin{figure}[!htbp]
\begin{center}
\includegraphics[angle=0,scale=0.60]{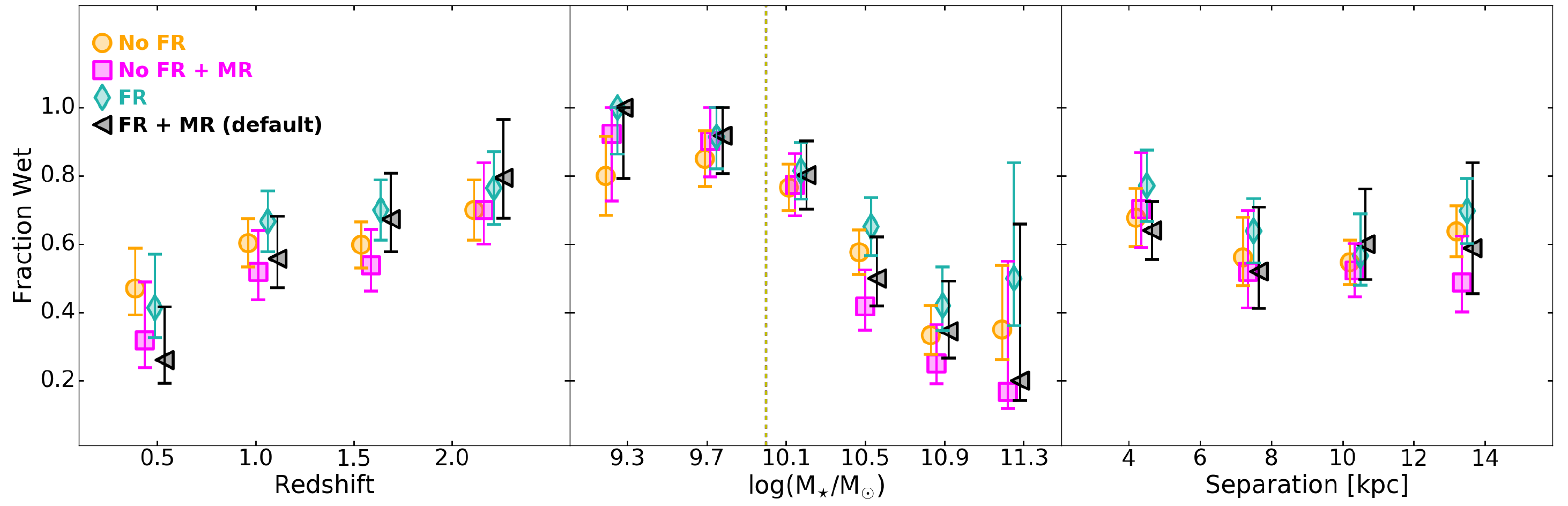}
\caption{ Fraction     of wet mergers as a function of redshift, stellar mass, and separation for merging galaxies selected using different flux and mass  ratios indicated in Figure \ref{fracq_z_cratios}. 
 \label{frac_wet_cratio}}
\end{center}
\end{figure}

Figure \ref{frac_wet_cratio} shows the comparison for the fraction of wet mergers as a function of redshift, stellar mass, and separation for the different ratio cuts. In all these plots the fraction of galaxies in wet mergers are within the uncertainties of the different results when different ratio cuts are applied.  For the different selection, the trends are all consistent. 

\begin{figure}[!htbp]
\begin{center}
\includegraphics[angle=0,scale=0.6]{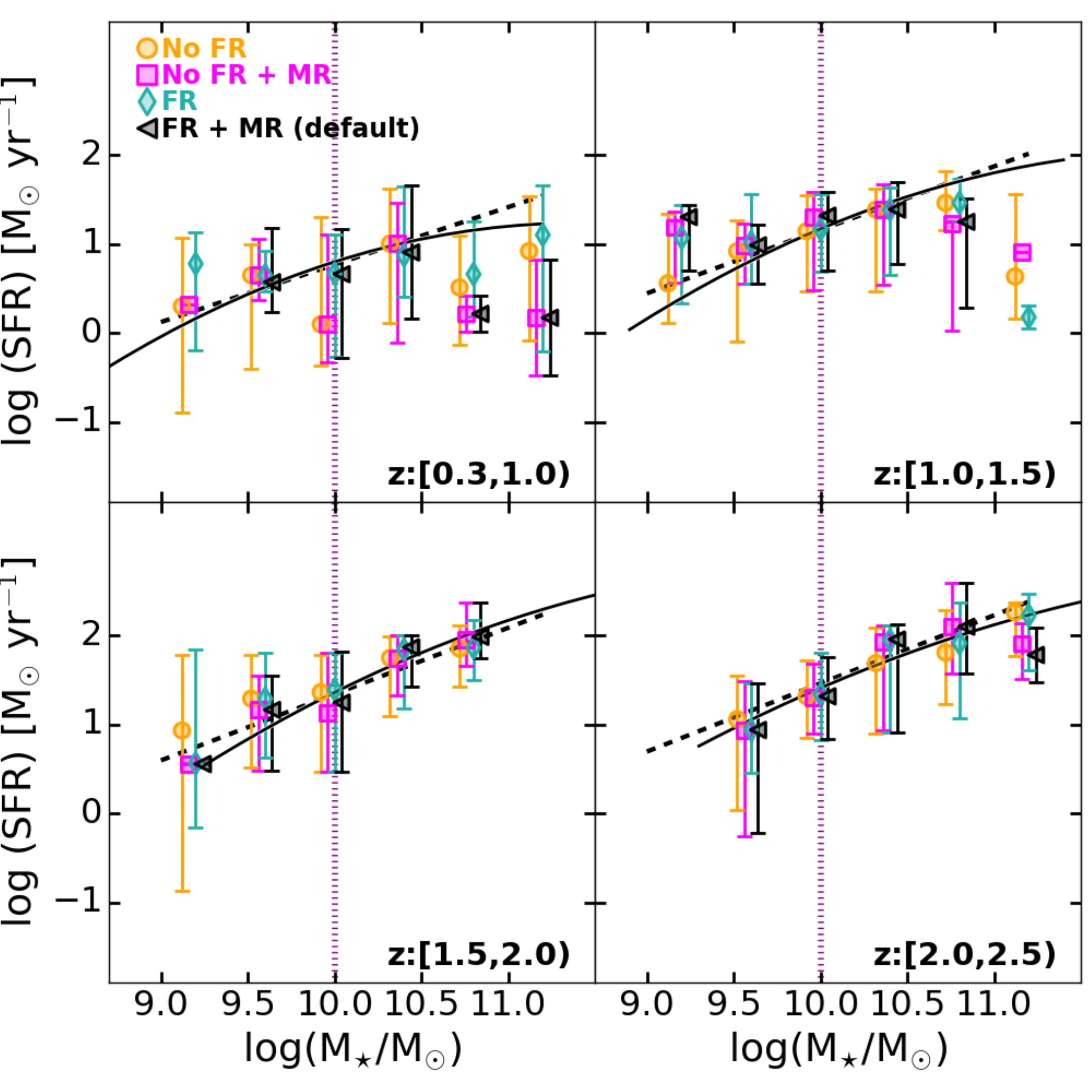}
\caption{ Star formation rate as a function of stellar mass for merging galaxies selected using different flux and mass ratios. 
 \label{mseq_cratio}}
\end{center}
\end{figure}

Finally, Figure \ref{mseq_cratio}  also shows consistent results between the different selection restrictions, indicating that  overall the mass and flux ratio cuts only affect quantitatively the number of selected mergers, but their  overall properties remain the same. 

\section{Impact of using a single band to select mergers} \label{comparison_bands}

We run the peak-finding algorithm on I814 stamps for galaxies at $0.3<z<1.0$ in the CANDELS fields, with the exception of the GOODS-North field, where we have only  the I850-band available.  Because we want to select merging galaxies redward of the 4000$\AA$  break, the I814 band is limited to $z<1.0$.
We want to probe that the selection of merging galaxies using the I814 and the H160 independently do not affect the results shown in this paper. 
In this selection we keep galaxies with $\log(\rm M_{\star}/M_{\odot})<10$.

\begin{figure}[!htbp]
\begin{center}
\includegraphics[angle=0,scale=0.4]{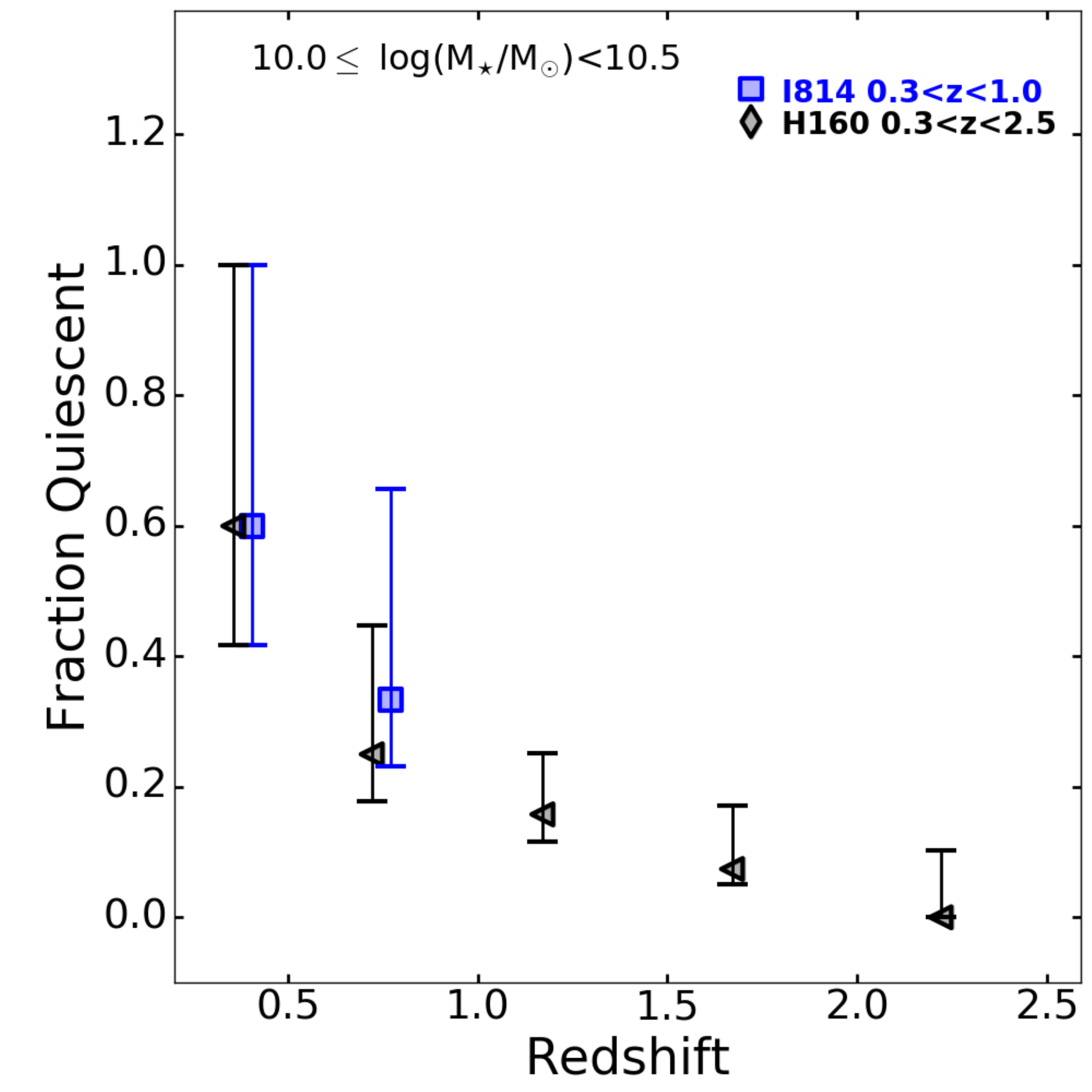}
\includegraphics[angle=0,scale=0.4]{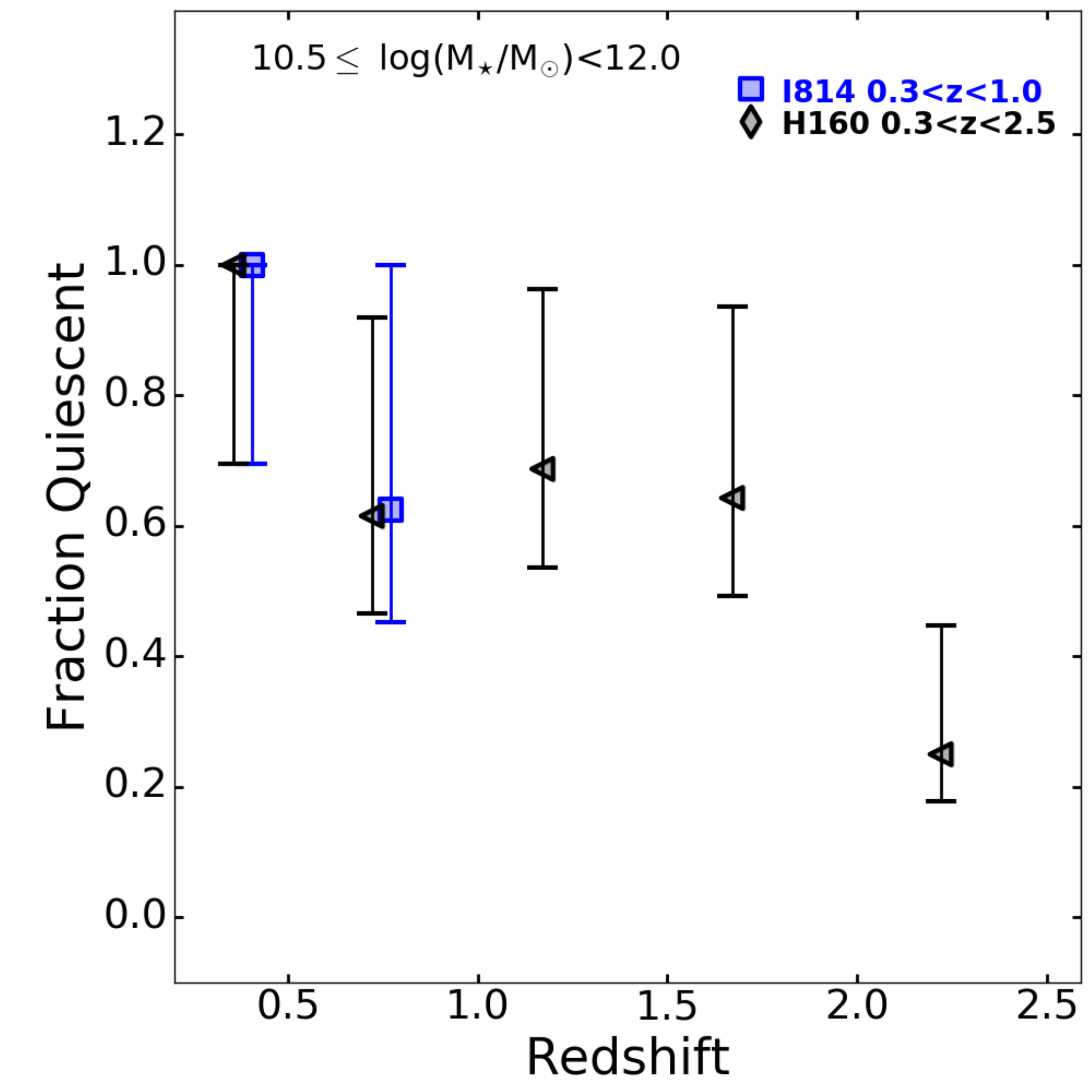}
\caption{ Fraction of quiescent galaxies as a function of redshift for galaxies in mergers with masses $10.0\le \log(\rm M_{\star}/M_{\odot})<10.5$ (left) and $10.5\le \log(\rm M_{\star}/M_{\odot})<12$ (right). The black diamonds shows the selection using the H160-band in the redshift range $0.3<z<2.5$, while the blue squares is for merging galaxies at $0.3<z<1.0$ obtained using the I814-band images. 
 \label{fracq_z_comp_bands}}
\end{center}
\end{figure}

\begin{figure}[!htbp]
\begin{center}
\includegraphics[angle=0,scale=0.4]{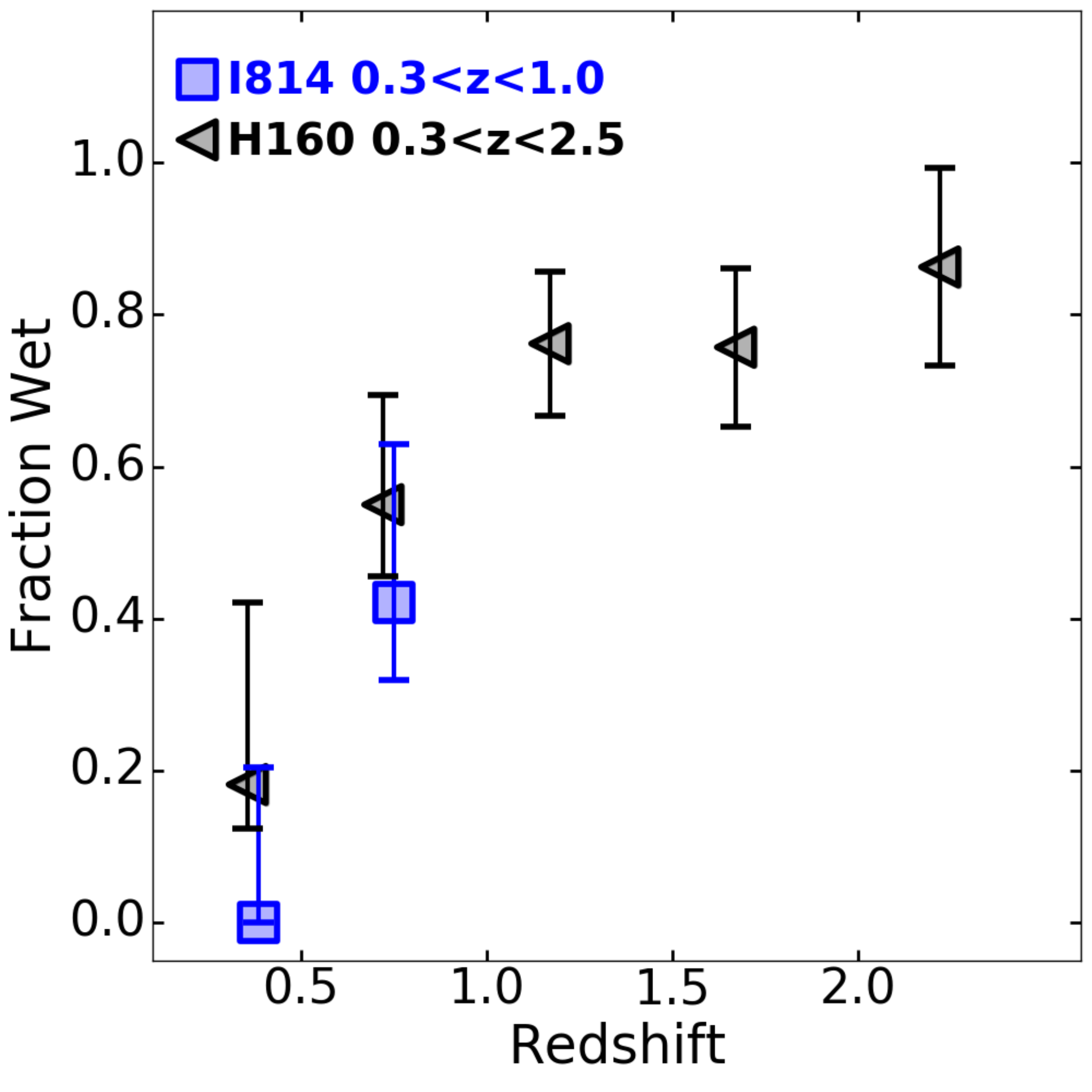}
\includegraphics[angle=0,scale=0.4]{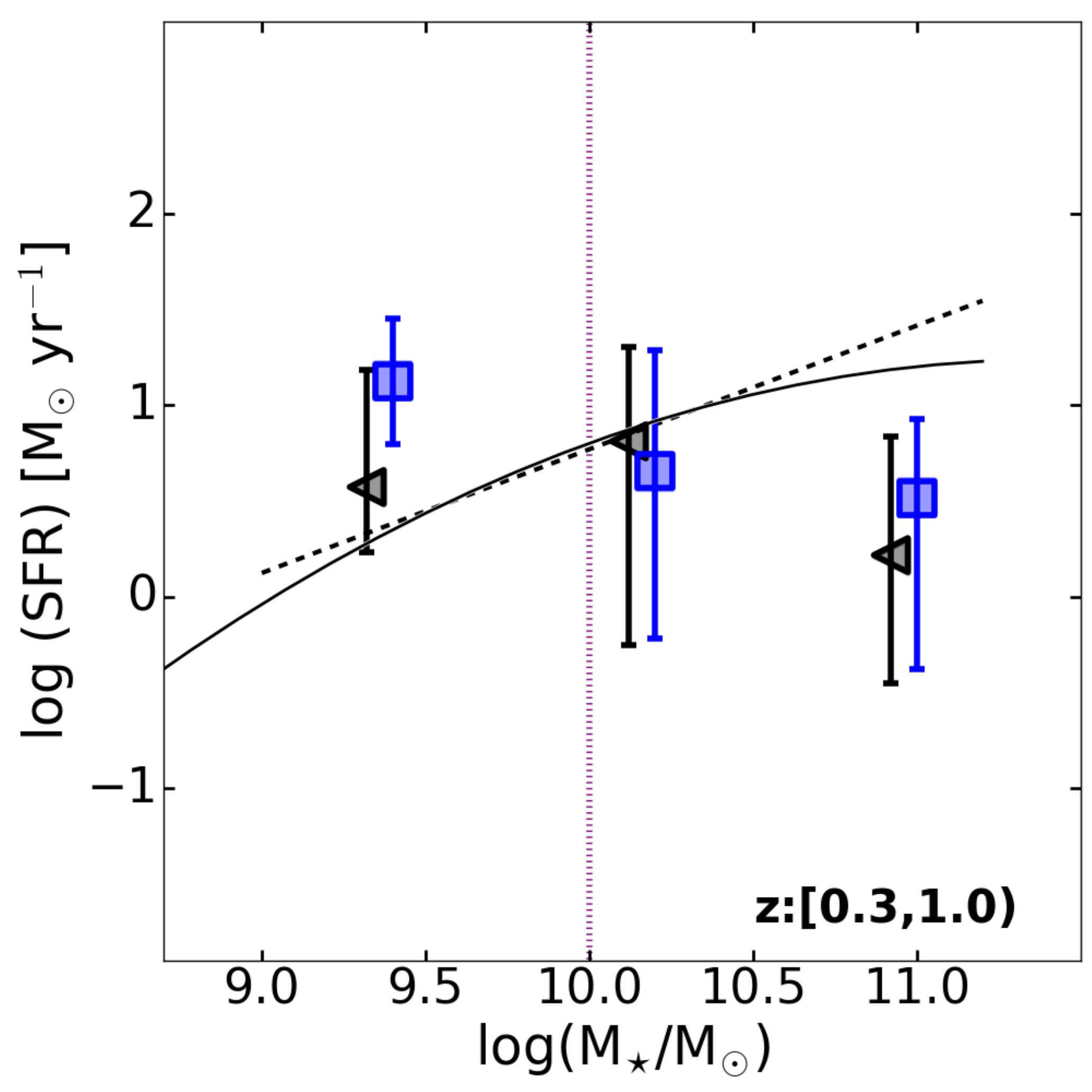}
\caption{ Fraction     of wet mergers as a function of redshift  for galaxies selected using the H160 and the I814 band (left). 
Right plot shows that median of the star formation rate as a function of stellar mass for these two samples.  
Lower and upper error bars correspond to the 15 and 85 percentiles of the distribution.  The dashed and solid curves
 correspond to the star formation  main sequence fits presented in \citet{speagle2014} and \citet{whitaker2014}, respectively. 
 \label{frac_wet_comp_bands}}
\end{center}
\end{figure}

Figure \ref{fracq_z_comp_bands} shows the fraction of quiescent galaxies as a function of redshift in two bins of mass, while Figure  \ref{frac_wet_comp_bands} presents the fraction of wet mergers as a function of redshift, and the median of the star formation rate as a function of stellar mass  for galaxies selected using the I814 and the H160 band. In all these figures we find no significant difference between the results.

\end{appendix}



\begin{thebibliography}{}

\bibitem[Armus et al.(1987)]{armus1987} Armus, L., Heckman, T., \& Miley, G.\ 1987, \aj, 94, 831 
\bibitem[Barnes \& Hernquist(1996)]{barnes1996} Barnes, J.~E., \& Hernquist, L.\ 1996, \apj, 471, 115 
\bibitem[B{\'e}thermin et al.(2015)]{bethermin2015} B{\'e}thermin, M., Daddi, E., Magdis, G., et al.\ 2015, \aap, 573, A113 
\bibitem[Brammer et al.(2008)]{brammer2008} Brammer, G.~B., van Dokkum, P.~G., \& Coppi, P.\ 2008, \apj, 686, 1503-1513 
\bibitem[Bridge et al.(2010)]{bridge2010} Bridge, C.~R., Carlberg, R.~G., \& Sullivan, M.\ 2010, \apj, 709, 1067 
\bibitem[Bruzual \& Charlot(2003)]{bruzual2003} Bruzual, G., \& Charlot, S.\ 2003, \mnras, 344, 1000 
\bibitem[Cattaneo et al.(2008)]{cattaneo2008} Cattaneo, A., Dekel, A., Faber, S.~M., \& Guiderdoni, B.\ 2008, \mnras, 389, 567 
\bibitem[Chabrier(2003)]{chabrier2003} Chabrier, G.\ 2003, \pasp, 115, 763 
\bibitem[Cibinel et al.(2015)]{cibinel2015} Cibinel, A., Le Floc'h, E., Perret, V., et al.\ 2015, \apj, 805, 181 
\bibitem[Cisternas et al.(2011)]{cisternas2011} Cisternas, M., Jahnke, K., Inskip, K.~J., et al.\ 2011, \apj, 726, 57 
\bibitem[Cole et al.(2008)]{cole2008} Cole, S., Helly, J., Frenk, C.~S., \& Parkinson, H.\ 2008, \mnras, 383, 546 
\bibitem[Cox et al.(2008)]{cox2008} Cox, T.~J., Jonsson, P., Somerville, R.~S., Primack, J.~R., \& Dekel, A.\ 2008, \mnras, 384, 386 
\bibitem[Daddi et al.(2010)]{daddi2010} Daddi, E., Bournaud, F., Walter, F., et al.\ 2010, \apj, 713, 686 
\bibitem[Davies et al.(2015)]{davies2015} Davies, L.~J.~M., Robotham, A.~S.~G., Driver, S.~P., et al.\ 2015, \mnras, 452, 616 
\bibitem[Dekel et al.(2009)]{dekel2009} Dekel, A., Birnboim, Y., Engel, G., et al.\ 2009, \nat, 457, 451 
\bibitem[Di Matteo et al.(2007)]{dimatteo2007} Di Matteo, P., Combes, F., Melchior, A.-L., \& Semelin, B.\ 2007, \aap, 468, 61 
\bibitem[Ellison et al.(2008)]{ellison2008} Ellison, S.~L., Patton, D.~R., Simard, L., \& McConnachie, A.~W.\ 2008, \aj, 135, 1877 
\bibitem[Ellison et al.(2013)]{ellison2013} Ellison, S.~L., Mendel, J.~T., Scudder, J.~M., Patton, D.~R., \& Palmer, M.~J.~D.\ 2013, \mnras, 430, 3128
\bibitem[Fensch et al.(2017)]{fensch2017} Fensch, J., Renaud, F., Bournaud, F., et al.\ 2017, \mnras, 465, 1934 
\bibitem[Gehrels(1986)]{gehrels1986} Gehrels, N.\ 1986, \apj, 303, 336 
\bibitem[Guo et al.(2012)]{guo2012} Guo, Y., Giavalisco, M., Ferguson, H.~C., Cassata, P., \& Koekemoer, A.~M.\ 2012, \apj, 757, 120 
\bibitem[Hopkins et al.(2006)]{hopkins2006} Hopkins, P.~F., Hernquist, L., Cox, T.~J., et al.\ 2006, \apjs, 163, 1 
\bibitem[Hopkins et al.(2008)]{hopkins2008} Hopkins, P.~F., Hernquist, L., Cox, T.~J., \& Kere{\v s}, D.\ 2008, \apjs, 175, 356-389 
\bibitem[Hung et al.(2013)]{hung2013} Hung, C.-L., Sanders, D.~B., Casey, C.~M., et al.\ 2013, \apj, 778, 129 
\bibitem[Kartaltepe et al.(2010)]{kartaltepe2010} Kartaltepe, J.~S., Sanders, D.~B., Le Floc'h, E., et al.\ 2010, \apj, 721, 98 
\bibitem[Kartaltepe et al.(2012)]{kartaltepe2012} Kartaltepe, J.~S., Dickinson, M., Alexander, D.~M., et al.\ 2012, \apj, 757, 23 
\bibitem[Kartaltepe et al.(2015)]{kartaltepe2015} Kartaltepe, J.~S., Mozena, M., Kocevski, D., et al.\ 2015, \apjs, 221, 11 
\bibitem[Kauffmann \& White(1993)]{kauffmann1993} Kauffmann, G., \& White, S.~D.~M.\ 1993, \mnras, 261
\bibitem[Kauffmann et al.(1999)]{kauffmann1999} Kauffmann, G., Colberg, J.~M., Diaferio, A., \& White, S.~D.~M.\ 1999, \mnras, 303, 188 
\bibitem[Kaviraj et al.(2013)]{kaviraj2013} Kaviraj, S., Cohen, S., Windhorst, R.~A., et al.\ 2013, \mnras, 429, L40 
\bibitem[Kaviraj et al.(2015)]{kaviraj2015} Kaviraj, S., Devriendt, J., Dubois, Y., et al.\ 2015, \mnras, 452, 2845 
\bibitem[Khochfar \& Burkert(2003)]{khochfar2003} Khochfar, S., \& Burkert, A.\ 2003, \apjl, 597, L117 
\bibitem[Kriek et al.(2009)]{kriek2009} Kriek, M., van Dokkum, P.~G., Labb{\'e}, I., et al.\ 2009, \apj, 700, 221 
\bibitem[Lackner et al.(2014)]{lackner2014} Lackner, C.~N., Silverman, J.~D., Salvato, M., et al.\ 2014, \aj, 148, 137 
\bibitem[Lanz et al.(2013)]{lanz2013} Lanz, L., Zezas, A., Brassington, N., et al.\ 2013, \apj, 768, 90
\bibitem[Li et al.(2008)]{li2008} Li, C., Kauffmann, G., Heckman, T.~M., Jing, Y.~P., \& White, S.~D.~M.\ 2008, \mnras, 385, 1903 
\bibitem[Lin et al.(2008)]{lin2008} Lin, L., Patton, D.~R., Koo, D.~C., et al.\ 2008, \apj, 681, 232-243 
 \bibitem[Lin et al.(2010)]{lin2010} Lin, L., Cooper, M.~C., Jian, H.-Y., et al.\ 2010, \apj, 718, 1158 
\bibitem[Lofthouse et al.(2017)]{lofthouse2017} Lofthouse, E.~K., Kaviraj, S., Conselice, C.~J., Mortlock, A., \& Hartley, W.\ 2017, \mnras, 465, 2895
\bibitem[Magdis et al.(2011)]{magdis2011} Magdis, G.~E., Daddi, E., Elbaz, D., et al.\ 2011, \apjl, 740, L15 
\bibitem[Man et al.(2016)]{man2016} Man, A.~W.~S., Zirm, A.~W., \& Toft, S.\ 2016, \apj, 830, 89 
\bibitem[Martis et al.(2016)]{martis2016} Martis, N.~S., Marchesini, D., Brammer, G.~B., et al.\ 2016, \apjl, 827, L25 
\bibitem[Mihos \& Hernquist(1996)]{mihos1996} Mihos, J.~C., \& Hernquist, L.\ 1996, \apj, 464, 641 
\bibitem[Momcheva et al.(2016)]{momcheva2016} Momcheva, I.~G., Brammer, G.~B., van Dokkum, P.~G., et al.\ 2016, \apjs, 225, 27 
\bibitem[Mundy et al.(2017)]{mundy2017} Mundy, C.~J., Conselice, C.~J., Duncan, K.~J., et al.\ 2017, \mnras, 470, 3507 
\bibitem[Neistein \& Dekel(2008)]{neistein2008} Neistein, E., \& Dekel, A.\ 2008, \mnras, 388, 1792 
\bibitem[Parkinson et al.(2012)]{parkinson2012} Parkinson, D., Riemer-S{\o}rensen, S., Blake, C., et al.\ 2012, \prd, 86, 103518 
\bibitem[Patton et al.(2013)]{patton2013} Patton, D.~R., Torrey, P., Ellison, S.~L., Mendel, J.~T., \& Scudder, J.~M.\ 2013, \mnras, 433, L59 
\bibitem[Peng et al.(2002)]{peng2002} Peng, C.~Y., Ho, L.~C., Impey, C.~D., \& Rix, H.-W.\ 2002, \aj, 124, 266 
\bibitem[Perlmutter et al.(1999)]{perlmutter1999} Perlmutter, S., Aldering, G., Goldhaber, G., et al.\ 1999, \apj, 517, 565 
\bibitem[Perret et al.(2014)]{perret2014} Perret, V., Renaud, F., Epinat, B., et al.\ 2014, \aap, 562, A1 
\bibitem[Puech et al.(2014)]{puech2014} Puech, M., Hammer, F., Rodrigues, M., et al.\ 2014, \mnras, 443, L49 
\bibitem[Riess et al.(1998)]{riess1998} Riess, A.~G., Filippenko, A.~V., Challis, P., et al.\ 1998, \aj, 116, 1009 
\bibitem[Robaina et al.(2010)]{robaina2010} Robaina, A.~R., Bell, E.~F., van der Wel, A., et al.\ 2010, \apj, 719, 844 
\bibitem[Rodighiero et al.(2011)]{rodighiero2011} Rodighiero, G., Daddi, E., Baronchelli, I., et al.\ 2011, \apjl, 739, L40 
\bibitem[Sanders \& Mirabel(1996)]{sanders1996} Sanders, D.~B., \& Mirabel, I.~F.\ 1996, \araa, 34, 749 
\bibitem[Scoville et al.(2014)]{scoville2014} Scoville, N., Aussel, H., Sheth, K., et al.\ 2014, \apj, 783, 84 
\bibitem[Scoville et al.(2016)]{scoville2016} Scoville, N., Sheth, K., Aussel, H., et al.\ 2016, \apj, 820, 83 
\bibitem[Scudder et al.(2015)]{scudder2015} Scudder, J.~M., Ellison, S.~L., Momjian, E., et al.\ 2015, \mnras, 449, 3719 
\bibitem[Silverman et al.(2015)]{silverman2015} Silverman, J.~D., Daddi, E., Rodighiero, G., et al.\ 2015, \apjl, 812, L23 
\bibitem[Skelton et al.(2014)]{skelton2014} Skelton, R.~E., Whitaker, K.~E., Momcheva, I.~G., et al.\ 2014, \apjs, 214, 24 
\bibitem[Speagle et al.(2014)]{speagle2014} Speagle, J.~S., Steinhardt, C.~L., Capak, P.~L., \& Silverman, J.~D.\ 2014, \apjs, 214, 15 
ons, B.~D., Urry, C.~M., Schawinski, K., Cardamone, C., \& Glikman, E.\ 2012, \apj, 761, 75 
\bibitem[Springel et al.(2005)]{springel2005} Springel, V., Di Matteo, T., \& Hernquist, L.\ 2005, \apjl, 620, L79 
\bibitem[Tacconi et al.(2010)]{tacconi2010} Tacconi, L.~J., Genzel, R., Neri, R., et al.\ 2010, \nat, 463, 781 
\bibitem[Tasca et al.(2014)]{tasca2014} Tasca, L.~A.~M., Le F{\`e}vre, O., L{\'o}pez-Sanjuan, C., et al.\ 2014, \aap, 565, A10 
\bibitem[Whitaker et al.(2012)]{whitaker2012} Whitaker, K.~E., van Dokkum, P.~G., Brammer, G., \& Franx, M.\ 2012, \apjl, 754, L29 
\bibitem[Whitaker et al.(2014)]{whitaker2014} Whitaker, K.~E., Franx, M., Leja, J., et al.\ 2014, \apj, 795, 104 
\bibitem[Whitaker et al.(2015)]{whitaker2015} Whitaker, K.~E., Franx, M., Bezanson, R., et al.\ 2015, \apjl, 811, L12 
\bibitem[White \& Frenk(1991)]{white1991} White, S.~D.~M., \& Frenk, C.~S.\ 1991, \apj, 379, 52 
\bibitem[Williams et al.(2011)]{williams2011} Williams, R.~J., Quadri, R.~F., \& Franx, M.\ 2011, \apjl, 738, L25 
\bibitem[Woods et al.(2010)]{woods2010} Woods, D.~F., Geller, M.~J., Kurtz, M.~J., et al.\ 2010, \aj, 139, 1857 
\bibitem[Wuyts et al.(2012)]{wuyts2012} Wuyts, S., F{\"o}rster Schreiber, N.~M., Genzel, R., et al.\ 2012, \apj, 753, 114 
\bibitem[Yuan et al.(2012)]{yuan2012} Yuan, F.-T., Takeuchi, T.~T., Matsuoka, Y., et al.\ 2012, \aap, 548, A117 

\end{thebibliography}
\end{document}